\documentclass[aps,pra,twocolumn,amsmath,amssymb, superscriptaddress,tightenlines]{revtex4}
\usepackage{graphicx}
\usepackage{epsfig}
\usepackage{dcolumn}
\usepackage{bm}
\usepackage{amssymb}
\usepackage{bm}
\usepackage{amsmath, bm}
\usepackage{upgreek}
\usepackage[colorlinks,citecolor=blue,linkcolor=blue,hyperindex]{hyperref}

\begin{document}

\title{Identification of topological order in the fractional quantum Hall state at $\nu=1/4$}
\author{Ken K. W. Ma}
\affiliation{Department of Physics, Brown University, Providence, Rhode Island 02912, USA}
\date{\today}


\begin{abstract}

The nature of the fractional quantum Hall state at quarter filling in a wide quantum well is still under debate. Both one-component non-Abelian and two-component Abelian orders have been proposed to describe the system. Interestingly, these candidates received support from different experiments under disparate conditions. In this article, we focus on non-Abelian orders from Cooper pairing between composite fermions and the Abelian Halperin-(5,5,3) order. We discuss and predict systematically different experimental signatures to identify them in future experiments. In particular, we address the Mach-Zehnder interferometry experiment and show that it can identify the recently proposed 22111 parton order.

\end{abstract}

\maketitle


\section{Introduction} \label{sec:intro}

The majority of the incompressible fractional quantum Hall (FQH) states in a two-dimensional electron gas have odd denominators in their filling factors, which can be well explained by the Haldane-Halperin hierarchy~\cite{Haldane_hierarchy, Halperin_hierarchy} and the theory of composite fermion~\cite{Jain_book}. In particular, the latter theory attaches $2p$  fluxes to an electron, such that the composite fermions experience a reduced average effective magnetic field. Furthermore, this magnetic field vanishes when the filling factor of the electron gas attains 
$\nu=n+1/2p$. Thus the system is expected to be gapless~\cite{HLR_theory}. This picture received experimental support from the observation of a well-defined Fermi sea of composite fermions in geometric resonance measurements at $\nu=5/2$~\cite{dima:W-CF, polarized-CF} and $\nu=1/4$~\cite{4-CF-exp}. At the same time, quantum Hall plateaus were observed at $\nu=5/2$~\cite{Willet1987, Pan1999} which led to a great surprise to the condensed matter society. Nowadays, more half-integer FQH states have been observed in different systems, such as ZnO heterostructures~\cite{Falson2015, Falson2018} and graphene-based devices~\cite{221_graphene_exp, Li2017, Zibrov2018}. In order to explain the incompressible FQH states with even-denominator filling factors, the idea of superconducting pairing between composite fermions was introduced~\cite{Read-Green, Scarola_nature}. Numerous topological orders have been proposed as candidates for the $\nu=5/2$ FQH state~\cite{MR1991, APf_Lee2007, APf_Levin2007, Son2015, Zucker2016, Jain-221, Jain-221-2, Halperin331, Balram-APf2018, Guang2013, Guang2014, Wen_Zee}. As a result, the nature of the FQH state is still under debate. In particular, some of these orders host non-Abelian quasiparticle excitations, which may open the door to topological quantum computation~\cite{Kitaev2003, Das-Sarma}. 

Apart from $\nu=5/2$, the FQH state was observed at $\nu=1/4$ in wide GaAs quantum wells~\cite{exp-wide-well1, exp-wide-well2, exp-wide-well3} and monolayer graphene at the isospin transition point~\cite{Zibrov2018}. In a wide quantum well, electrons tend to minimize the energy by concentrating themselves near to the two sides of the well. This charge distribution leads to an effective bilayer system~\cite{Suen1991, Suen1994, Papic-1/4}. Using the language of pseudospin, one may associate the spin-up and spin-down states to the two lowest electronic subbands of the system. This additional degree of freedom allows the formation of the $\nu=1/4$ FQH state~\cite{Papic-1/4, Papic-Regnault-DS}. The two subbands are separated by a gap $\Delta_{\rm SAS}$, which depends on both the width of the well and the electron density of the system. In principle, both one-component and two-component topological orders can be realized in a bilayer system. Which one is preferred depends on the competition between $\Delta_{\rm SAS}$ and the interaction between electrons in each effective layer, $e^2/\left(\epsilon \ell_B\right)$. With a typical width of the quantum well $w\approx (50-60)$ nm and an electron density $n\approx (2.0-2.6)\times 10^{11}$ cm$^{-2}$ in the experiment~\cite{exp-wide-well1, exp-wide-well2, exp-wide-well3}, it was estimated that $\Delta_{\rm SAS}/(e^2/\epsilon \ell_B)\lesssim 0.1$~\cite{Papic-1/4, wide-well}.

Similar to the case of $\nu=5/2$ FQHE, different topological orders have been proposed to describe the $\nu=1/4$ FQH state~\cite{MacDonald1990, Papic-1/4, wide-well, Barkeshli2010-1, Barkeshli2010-2, Lu2010, Scarola-PRB2010}. The original experiment by Luhman \textit{et al.}~\cite{exp-wide-well1} reported that the FQH state was strengthened by tilting the sample in a magnetic field. Since it is believed that $\Delta_{\rm SAS}$ is reduced by the in-plane magnetic field~\cite{Hu1992, Lay1997}, the experiment was interpreted to favor a two-component topological order in the system. By investigating the problem numerically, Papi\'{c} \textit{et al.}~\cite{Papic-1/4} concluded that there is a competition between the Abelian Halperin-(5,5,3) order and the non-Abelian Pfaffian order in the system being explored in Ref.~\cite{exp-wide-well1}. At the same time, they pointed out that the two-component state might be further stabilized by the in-plane magnetic field applied in the experiment. 

The effect of charge distribution on the $\nu=1/4$ FQH state was examined in later experiment~\cite{exp-wide-well2, exp-wide-well3}. On the one hand, Ref.~\cite{exp-wide-well2} reported that the FQH state disappeared when the charge density is lowered or the charge distribution was made asymmetric. This result supported the Halperin-(5,5,3) order. On the other hand, observation of the FQH state in a sample with highly asymmetric charge distribution, and its disappearance when the distribution became symmetric, seemingly favored the one-component Pfaffian state~\cite{exp-wide-well3}. Later on, an alternative explanation to the result in Ref.~\cite{exp-wide-well3} with an Abelian two-component state based on partial subband polarization was proposed~\cite{Scarola-PRB2010}.

Very recently, Faugno and his collaborators have reexamined the phase diagram of the quantum well problem at $\nu=1/4$~\cite{wide-well}. Their numerical results suggested the possibility of realizing a 22111 parton order in the system. Different from previous proposals, the 22111 parton order is topologically equivalent to a paired state formed by Cooper pairing of composite fermions in the $f$-wave channel. At this stage, the nature of the $\nu=1/4$ FQH state observed in the wide quantum well remains unsettled. In fact, the interplay between interlayer tunneling, charge imbalance, and the nature of quantum Hall state in a wide quantum well can be complicated~\cite{Peterson-Papic-DS, Scarola-PRB2010, Thiebaut2014}. Given that the details and the procedures in different experiments were quite disparate, it may be possible that different topological orders were realized in different cases. Therefore, it is important to have a detailed list of predicted experimental signatures to identify different topological orders in future experiments. This is the main motivation of our current paper.

In our previous work~\cite{16-fold}, we have related different topological orders for half-integer FQHE and two-dimensional topological superconductors built by composite fermions. Based on this connection, we have systematically classified the orders by Kitaev's sixteenfold way~\cite{Kitaev} and predicted their signatures in different experiments. In this paper, we continue our work in this direction to study different non-Abelian orders for the FQH state at $\nu=1/4$. At the same time, it is equally important to understand the experimental signatures for Abelian orders. Since the Halperin-(5,5,3) order was shown to be a leading candidate in this category~\cite{Papic-1/4}, we will examine it explicitly. From the results in this paper, we argue that different topological orders can be identified unambiguously by combining signatures from various experiments. In turn, the question of whether a one-component or a two-component order is realized in the wide quantum well system under different conditions may be answered.

The paper is organized as follows. First, we discuss the topological properties of different non-Abelian orders for $\nu=1/4$ FQHE and make predictions on their experimental signatures in Sec.~\ref{sec:wavefunction-NA}. Then, we examine systematically the tunneling current and Fano factor in Mach-Zehnder interferometry for each non-Abelian topological order in Sec.~\ref{sec:MZ-exp}. In particular, we discuss how the recent proposal on 22111 parton order can be tested by the Mach-Zehnder inteferometry experiment. In Sec.~\ref{sec:553}, we provide a discussion on the experimental signatures for the Abelian two-component Halperin-(5,5,3) order. The results in the previous three sections are summarized in Sec.~\ref{sec:summary-exp}. In the same section, we briefly comment on how the nature of $\nu=1/4$ FQHE in the wide quantum well can be resolved by combining different experimental signatures. Finally, we conclude our work in Sec.~\ref{sec:conclusion}. At the end of the paper, three Appendixes are provided to supplement the main text. Appendix~\ref{app:Chern} provides an explicit calculation on the Chern number for a chiral $l$-wave paired state. Appendix~\ref{app:wave function} introduces a class of simple wave functions for non-Abelian orders by solving the Bardeen-Cooper-Schrieffer Hamiltonian. Appendix~\ref{app:NA-2C} provides a brief discussion of several other two-component candidates for fractional quantum Hall state at $\nu=1/4$.

\section{One-component non-Abelian orders for $\nu=1/4$ FQHE} 
\label{sec:wavefunction-NA}

In this section, we focus on one-component non-Abelian orders for the $\nu=1/4$ FQH state originating from Cooper pairing between spin-polarized composite fermions. The pairing is described by the following mean-field Bardeen-Cooper-Schrieffer (BCS) Hamiltonian:
\begin{eqnarray} \label{eq:BCS_hamiltonian}
H_{\rm BCS}
=\sum_{\bm k} \left[
\xi_{\bm k}c^\dagger_{\bm k}c_{\bm k}
+\frac{1}{2}\left(\Delta_{\bm k}^* c_{-\bm k}c_{\bm k}
+\Delta_{\bm k}c^\dagger_{\bm k}c^\dagger_{-\bm k}\right)
\right].
\end{eqnarray}
In the above equation, $\xi_{\bm k}=k^2/2m-\mu$, with $m$ and $\mu$ being the effective mass and the chemical potential of the composite fermions, respectively. Also, we set $\hbar=1$ throughout the paper. The symbol $\Delta_{\bm k}$ denotes the pairing gap function. In this paper, we focus on the chiral $l$-wave pairing, such that $\Delta_{\bm k}=\Delta_0\left(k_x\pm ik_y\right)^l$. Here, we need to clarify our notations. In the following discussion, $l$ is always positive. Meanwhile, we will also call the paired state with $\Delta_{\bm k}=\Delta_0\left(k_x- ik_y\right)^l$ the paired state with a negative $\ell$, where $\ell=-l$.

Since the composite fermions are spin-polarized, antisymmetry of the wave function only allows pairing in odd-$l$ channels. It was shown by Read and Green~\cite{Read-Green} that the system is in the weak pairing phase and exhibits nontrivial topology when $\mu>0$. In Appendix~\ref{app:Chern}, we evaluate the Chern number $\mathcal{C}$ for the bulk of the system exactly. It is found that $\mathcal{C}=\pm l$ for $\Delta_{\bm k}=\Delta_0\left(k_x\pm ik_y\right)^l$. In other words, there is a one-one correspondence between the Chern number and the pairing channel. Furthermore, the bulk-edge correspondence suggests that $l$ copropagating Majorana modes exist at the edge of the system. This conclusion agrees with the numerical result obtained in Ref.~\cite{Dubail_Read}. Note that a pair of Majorana modes can form a Dirac fermion. Furthermore, the statistics is Abelian if all edge modes are Dirac fermions. Given that $l$ is odd, there is at least one unpaired Majorana mode. Hence, the paired state with odd $l$ is described by a non-Abelian topological order~\cite{Read-Green}.

The wave function for Pfaffian order to the $\nu=1/4$ FQH state is given by
\begin{eqnarray} \label{eq:Pf_wavefunction}
\Psi_{\rm Pf}
=\text{Pf}\left(\frac{1}{z_i-z_j}\right)
\prod_{i<j}\left(z_i-z_j\right)^4.
\end{eqnarray}
Notice that the Gaussian exponential factor has been suppressed. Here, the fourth power in the Jastrow factor fixes the filling factor at $\nu=1/4$, which can also be understood as attaching four flux quanta to an electron and turning it into a composite fermion~\cite{Jain_book}. The Pfaffian factor originates from the BCS pairing between composite fermions in the $\ell=1$ channel. Following the procedures in Ref.~\cite{16-fold}, wave functions for other non-Abelian orders resulting from higher $l$-wave pairing can be constructed iteratively for the $\nu=1/4$ FQHE. A more detailed discussion on composite-fermion pairing and another class of wave functions for the $l$-wave paired state can be found in Appendix~\ref{app:wave function}.

The conformal field theory (CFT) approach provides a systematic way to extract topological properties of a topological order. It is conjectured that a wave function for a quantum Hall state can be constructed from correlation function between conformal field operators for electrons~\cite{Hansson-CFT}. For example, the Pfaffian wave function in Eq.~\eqref{eq:Pf_wavefunction} can be constructed from the following correlation function:
\begin{eqnarray} \label{eq:CFT-correlation}
\Big\langle \prod_k G_k(z_k)\Big\rangle
=\bigg\langle\prod_k
\psi(z_k)e^{4i\varphi_\rho(z_k)}
\bigg\rangle.
\end{eqnarray}
The complex variable $z_k=x_k +iy_k$ labels the positions of the electrons on the 2D plane. Here, 
$\psi$ is the Majorana mode, and $\varphi_\rho$ is the Bose charged mode. As a remark, an additional vertex operator to neutralize the background should also be included in $G_k$, which is not shown here.

\subsection{Quasiparticles and topological properties}
\label{sec:topo-property}

The edge structure of the $\ell$-wave paired state consists of two parts. First, it consists of $l$ chiral Majorana modes, $\psi_j$ with $j=1, 2, \cdots, l$. The corresponding Lagrangian density for them with the same velocity $v_n$ is:
\begin{eqnarray}
\mathcal{L}_\psi
=i\sum_{j=1}^l
\psi_j \left[\partial_t + v_n \text{sgn}\left(\ell\right)\partial_x\right]\psi_j.
\end{eqnarray}
Depending on the sign of $\ell$, these Majorana modes can be downstream or upstream. The second part is a single downstream charged mode $\varphi_\rho$ with velocity $v_\rho$, being described by the following Lagrangian density:
\begin{eqnarray}
\mathcal{L}_\rho
=-\frac{4}{4\pi} 
\partial_x\varphi_\rho
\left(\partial_t\varphi_\rho+v_\rho \partial_x\varphi_\rho\right).
\end{eqnarray}

From the edge structure, one can write down the most relevant electron operator for the topological order as 
\begin{eqnarray}
\Psi_e=\psi_j e^{4i\varphi_\rho}.
\end{eqnarray}
At the same time, the operator product expansion (OPE) between a quasiparticle operator 
$\Psi_{\rm qp}$ and all possible electron operators must be single valued~\cite{Hansson-CFT}. Generically, we write $\Psi_{\rm qp} = \prod_j \sigma_j e^{i\omega\varphi_\rho}$. Here, $\sigma_j$ is the twist field with conformal dimension $h_\sigma=1/16$ in the SU(2)$_2$ CFT. Its fusion rule is
$\sigma_j\times\sigma_j=\psi_j+I$. The OPE between $\Psi_{\rm qp}$ with $\Psi_e$ gives
\begin{eqnarray}
\lim_{z\rightarrow w}
\left[\Psi_{\rm qp}(z)\Psi_{\rm e}(w)\right]
\sim \left(z-w\right)^{\omega-1/2}.
\end{eqnarray}
The single-valuedness condition leads to $\omega=1/2+n$, where $n$ is an integer. Thus the quasiparticle has charge 
\begin{eqnarray}
Q_{\rm qp}=\frac{e}{4}(n+1/2).
\end{eqnarray}
Therefore, the most fundamental quasiparticle has charge $e/8$.

\subsubsection{Fractional statistics}

From the edge structure and the form of quasiparticle operators, it is believed that the fractional statistics of the quasiparticles would satisfy the sixteenfold way. In fact, this is a universal feature of all paired states for FQHE at $\nu=1/2p$. It is because the non-Abelian sector, formed by the Majorana modes, is always described by the same CFT. Different filling factors of the FQH system correspond to different Abelian U(1) vertex operators for the charged mode only. 

For later discussion on Mach-Zehnder interferometry in Sec.~\ref{sec:MZ-exp}, we evaluate the phase accumulated when an $e/8$ quasiparticle makes a complete counterclockwise circle about another $e/8$ quasiparticle. There are two fusion channels for the non-Abelian neutral vortex 
$\sigma$ formed by the Majorana modes. Depending on the fusion channel $\beta=\psi$ or $I$, the phase accumulated is
\begin{align}
\label{eq:stat1}
\phi_\psi
&=\phi_{U(1)}
+\phi^{\sigma\sigma}_\psi
=\left(\frac{\pi}{8}+\frac{3\pi\ell}{4}\right)~(\text{mod}~2\pi),
\\
\label{eq:stat2}
\phi_I
&=\phi_{U(1)}
+\phi^{\sigma\sigma}_I
=\left(\frac{\pi}{8}-\frac{\pi\ell}{4}\right)~~(\text{mod}~2\pi).
\end{align}
The first term $\phi_{U(1)}=\pi/8$ comes from the Abelian U(1) sector, whereas the second term $\phi^{\sigma\sigma}_{\beta}$ comes from the braiding rules for neutral vortices in the sixteenfold way~\cite{16-fold, Kitaev}.

\subsubsection{Central charge and thermal Hall conductance}

With the edge structure discussed before, the central charge of the topological order can be determined easily. The single Bose mode and $l$ Majorana modes contribute $1$ and $l/2$ to the central charge, respectively. These two contributions add (subtract) when $\ell$ is positive (negative). Hence, the net central charge is
\begin{eqnarray}
c=1+\frac{\ell}{2}.
\end{eqnarray}

Existing thermal transport experiments cannot differentiate downstream modes and upstream modes~\cite{Banerjee2017, Banerjee2018}. Thus, a positive thermal Hall conductance $\kappa_H$ is measured. Furthermore, $\kappa_H$ depends on whether the edge of the system is thermally equilibrated or not. If the edge of the quantum Hall bar in the experiment is much longer than the thermal equilibration length, i.e., $L\gg\ell_{\rm th}$, then the edge is under full thermal equilibration. In this scenario, one has~\cite{Read-Green, Kane_thermal, Cappelli_thermal}
\begin{eqnarray} \label{eq:thermal-Hall}
\kappa_H
=\frac{\pi^2 k_B^2 T}{3h}\left| 1+\frac{\ell}{2} \right|.
\end{eqnarray}
This result is universal for all filling factors at $\nu=1/2p$.

\subsubsection{Scaling dimension and tunneling exponents}

Suppose quasiparticles can tunnel between two edges of the same FQH liquid in a tunneling experiment. It was predicted that the tunneling current and conductance satisfy the scaling laws: $I\sim V^{2g-1}$ and $G\sim T^{2g-2}$, respectively~\cite{Wen_book}. Here, $V$ is the voltage difference across the two edges and $T$ is the temperature of the system. The tunneling exponent $g$ is two times the scaling dimension of the quasiparticle operator. This exponent is universal for topological orders without upstream edge modes (i.e., pairing in 
$\ell>0$ channels). 

For $\ell<0$ (topological orders with upstream modes), the tunneling exponents are nonuniversal in a clean sample. Instead, they depend on the interaction between edge modes. On the other hand, impurities must exist in a real sample and lead to interedge tunneling. Suppose the disorder is weak and the corresponding interedge tunneling is a relevant process in the sense of renormalization group (RG). Then, the edge physics at low temperature is described by a disorder-dominated phase. In this case, we also say that the edge is equilibrated. Following the analysis in Refs.~\cite{APf_Levin2007} and~\cite{Guang2013}, one can conclude that the tunneling exponents are universal when $\ell\leq -3$. The exponents are also universal for the case with $\ell=-1$ (PH-Pfaffian order). It is because any random coupling between the charged mode and the single Majorana mode is irrelevant. 

For simplicity, we assume the edge is equilibrated by disorder throughout the paper. Under this assumption, the scaling dimensions for different types of quasiparticles are
\begin{eqnarray} \label{eq:scaling-dim}
\Delta_e=\frac{5}{2}
~,~
\Delta_{e/4}=\frac{1}{8}
~,~
\Delta_{e/8}=\frac{l}{16}+\frac{1}{32}.
\end{eqnarray}
Furthermore, suppose the tunneling process is dominated by $e/8$ quasiparticles (see the discussion in Sec.~\ref{sec:RG-analysis}). Then, one has
\begin{eqnarray} \label{eq:tun-exp}
g_{e/8}=2\Delta_{e/8}
=\frac{l}{8}+\frac{1}{16},
\end{eqnarray}
and the following scaling laws:
\begin{align} 
I&\sim V^{l/4-7/8},
\label{eq:I-1/4}
\\
G&\sim T^{l/4-15/8}.
\label{eq:G-1/4}
\end{align}
Equations~\eqref{eq:tun-exp}-\eqref{eq:G-1/4} may provide some information to identify the topological order in the $\nu=1/4$ FQHE from tunneling experiment.

\subsubsection{Shift and Hall viscosity}

In a numerical simulation, one may place a FQH liquid on a two-dimensional sphere. Since the sphere has a non-zero curvature, the number of magnetic flux quanta being enclosed 
(denoted as $N_\phi$) and the number of electrons $N$ are not simply related only by the filling factor. To quantify the difference from the plane geometry, the concept of shift $\mathcal{S}$ was defined as~\cite{Wen-Zee-shift}
\begin{eqnarray} \label{eq:def_shift}
N_\phi=N/\nu -\mathcal{S}.
\end{eqnarray}
It was shown that $\mathcal{S}$ is a topological quantum number, which depends on the topological order in the FQH liquid~\cite{Wen-Zee-shift}.

Since $\mathcal{S}$ is a topological number, its value for different non-Abelian orders can be found by examining the simpler form of wave functions in Eqs.~\eqref{eq:l>0} and~\eqref{eq:l<0}. We determine $\mathcal{S}$ by finding the highest power of $z_1$ in the wave function and treating $\bar{z}\sim 1/z$. This power is the same as $N_\phi$. For $\nu=1/2p=1/4$ and $\ell>0$, Eq.~\eqref{eq:l>0} gives
\begin{eqnarray}
N_\phi 
= 4(N-1)-l
= 4N-(4+l).
\end{eqnarray}
Similarly, one obtains from Eq.~\eqref{eq:l<0} for $\ell<0$
\begin{eqnarray} 
N_\phi 
= 4(N-1)+l
= 4N-(4-l).
\end{eqnarray} 
From the definition of $\mathcal{S}$ in Eq.~\eqref{eq:def_shift}, the $\ell$-wave paired state has
\begin{eqnarray} \label{eq:shift}
\mathcal{S}
=\ell+1/\nu
=\ell+4 .
\end{eqnarray}

Furthermore, the Hall viscosity is expected to be quantized as~\cite{Read-viscosity, Read-Rezayi-viscosity}:
\begin{eqnarray} \label{eq:Hall-viscosity}
\eta_H=\frac{\rho\mathcal{S}}{4}
=\frac{\rho}{4}\left(\ell +4\right),
\end{eqnarray}
where $\rho$ is the average electron density of the FQH liquid. Note that Eqs.~\eqref{eq:thermal-Hall}, ~\eqref{eq:tun-exp}, \eqref{eq:shift}, and~\eqref{eq:Hall-viscosity} agree with the results in Ref.~\cite{wide-well}.

\section{Mach-Zehnder Interferometry for $l$-wave paired state at $\nu=1/4$}
\label{sec:MZ-exp}

In this section, we build on the discussion in Sec.~\ref{sec:wavefunction-NA} to examine experimental signatures in Mach-Zehnder interferometry for each non-Abelian order. It is essential to remark that quasiparticle statistics depends on the topological properties of the topological order, but not the precise microscopic wave function. Indeed, different wave functions with the same topological properties can be formulated to describe the low energy physics of a quantum Hall system. Consequently, interferometry experiment may identify the topological nature of the state and decide if it is non-Abelian. However, it cannot determine the exact wave function of the system.

As we argued in Ref.~\cite{16-fold}, all non-Abelian topological orders in the sixteenfold way should demonstrate the even-odd effect in a Fabry-P\'{e}rot interferometer. As a result, this effect cannot help us to distinguish different non-Abelian orders for the $\nu=1/4$ FQH state. The ambiguity motivates us to examine a more complicated setup, namely the Mach-Zehnder interferometer.

\begin{figure}[htb]
\begin{center}
\includegraphics[width=3.0 in]{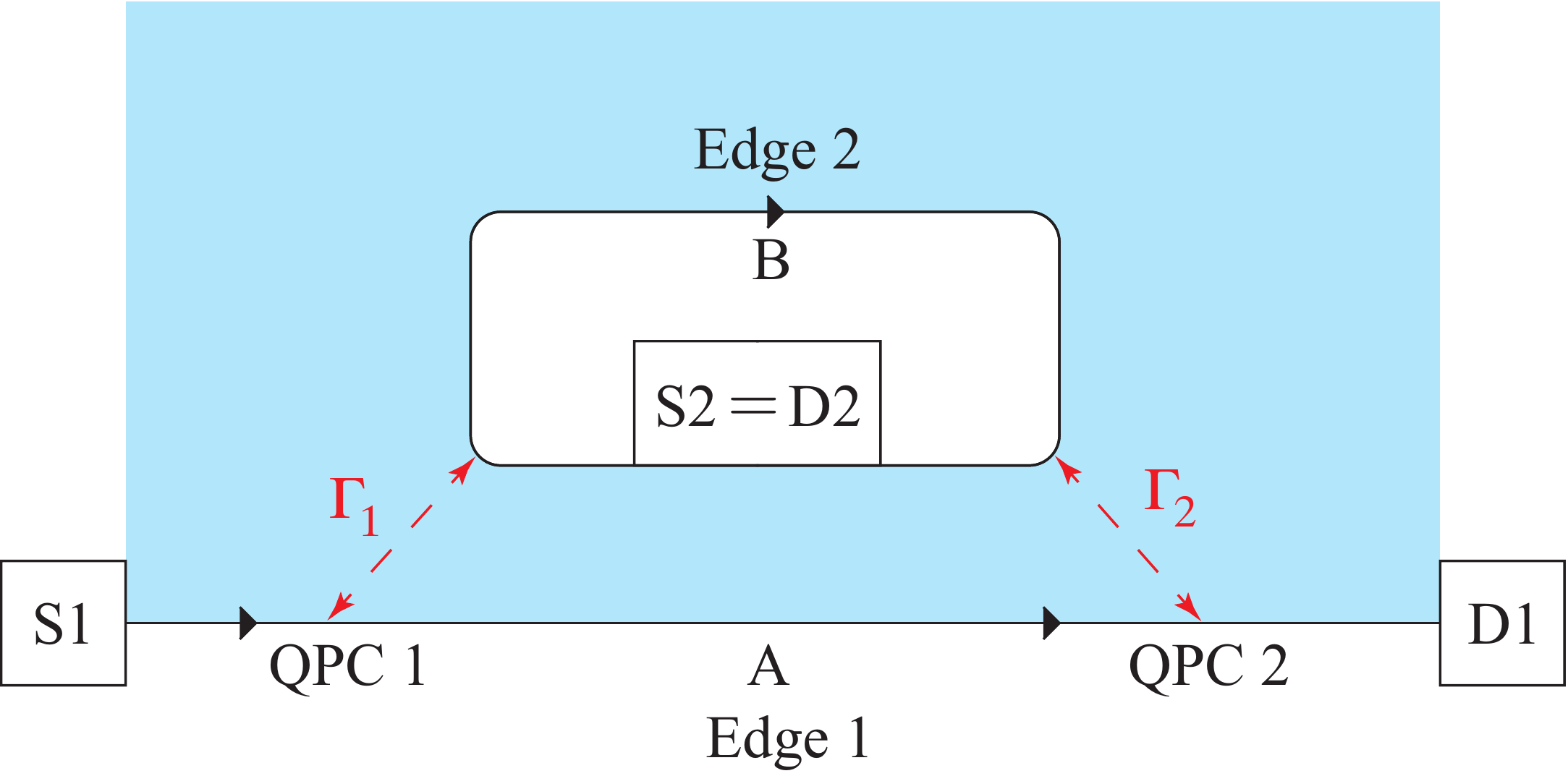}
\caption{Illustration of an electronic Mach-Zehnder interferometer. Charges propagate from
source S1 to drain D1 and source S2 to drain D2, as shown by the arrows in the figure. Quasiparticle tunneling between two edges is possible at the two quantum point contacts, namely QPC1 and QPC2. (Adopted from Ref.~\cite{16-fold}.)}
\label{fig:MZI}
\end{center}
\end{figure}

Now, we briefly review the principle of an electronic Mach-Zehnder interferometer. A more detailed discussion can be found in Refs.~\cite{KT2006, KT_noise, 16-fold}. A schematic plot for the interferometer is shown in Fig.~\ref{fig:MZI}. Quasiparticles can tunnel between the two edges of the quantum Hall liquid at the two quantum point contacts (QPCs), with tunneling amplitudes 
$\Gamma_1$ and $\Gamma_2$. The tunneling process is described by the following Hamiltonian:
\begin{eqnarray} \label{eq:H_tun}
H_{\rm tun}
=\int dx~\left[\Gamma_q \mathcal{O}_{\rm A}\mathcal{O}^\dagger_{\rm B}+ \text{H.c.}\right].
\end{eqnarray}
The symbol $\mathcal{O}_i$ denotes the operator for the charge-$q$ quasiparticles which tunnel at the QPC from edge $i=\rm{A}$ or $\rm{B}$, with tunneling amplitude $\Gamma_q$. Typically, a voltage $V$ is applied to S1 and leads to an electrochemical potential difference $eV$ between the two edges. In the experiment, the tunneling current from source S1 to drain D2 and the corresponding Fano noise are measured. Both quantities depend on the magnetic flux enclosed by the loop QPC1-A-QPC2-B-QPC1 and $V$. 

The state of D2 is described by a superselection sector in the form $(q,\alpha)$, where $q$ and $\alpha$ are the electric charge and topological charge being stored in D2, respectively. When both tunneling amplitudes at the two QPCs are small, and the fusion channel of the tunneling particle with the topological charge in D2 is known, the transition rate between two superselection sectors is given by~\cite{KT2006}:
\begin{align} \label{eq:transition-rate}
\nonumber
p\left(\phi_s\right)
=r\left[\left|\Gamma_1\right|^2 + \left|\Gamma_2\right|^2
+2\left|u\Gamma_1\Gamma_2\right|
\cos{\left(\phi_{\rm AB}+\phi_s+\delta\right)}\right].
\\
\end{align}
In the above equation, $\phi_{\rm AB}$ and $\phi_s$ are the Aharanov-Bohm phase and statistical phase accumulated when the quasiparticle encircles the whole device. Both constants $r$ and $u$ depend on the voltage and the temperature. The symbol $\delta$ is defined as $\delta=\text{arg}\left(u\Gamma_2/\Gamma_1\right)$.

\subsection{Renormalization group analysis}
\label{sec:RG-analysis}

In the present case, both charge-$e/8$ or charge-$e/4$ quasiparticles can tunnel at the QPCs. Thus one needs to identify which one of them dominates the process. The renormalization group equation for the tunneling process described by Eq.~\eqref{eq:H_tun} is
\begin{eqnarray} \label{eq:RG}
\frac{d\Gamma_q}{db}
=\left(1-2\Delta_q\right)\Gamma_q .
\end{eqnarray}
Here, $\Delta_q$ is the scaling dimension of the operator $\mathcal{O}_i$. From Eqs.~\eqref{eq:scaling-dim} and~\eqref{eq:RG}, one obtains the RG equations for $e/4$ and $e/8$ quasiparticles separately:
\begin{align}
\label{eq:RG-e/4}
\frac{d\Gamma_{e/4}}{db}
&=\frac{3}{4}\Gamma_{e/4},
\\
\label{eq:RG-e/8}
\frac{d\Gamma_{e/8}}{db}
&=\left(\frac{15}{16}-\frac{l}{8}\right)\Gamma_{e/8}.
\end{align}
Again, we remind that the edge is assumed to be equilibrated. Then, Eqs.~\eqref{eq:RG-e/4} and~\eqref{eq:RG-e/8} also hold for paired states with $\ell<0$. Aside from charged particles, a neutral fermion $\psi$ can tunnel at the QPC. The corresponding RG equation is
\begin{eqnarray}
\frac{d\Gamma_\psi}{db}
=\left(1-2\Delta_\psi\right)\Gamma_\psi
=0 .
\end{eqnarray}

From the above RG analysis, the tunneling process for the $e/4$ quasiparticle is always relevant. For $e/8$ quasiparticle tunneling, it is relevant when $l<7$. The neutral-fermion tunneling is marginally relevant~\cite{Fisher_Nayak2007}. Furthermore, one has $\Delta_{e/4}=1/8$ and 
$\Delta_{e/8}=l/16+1/32$ from Eq.~\eqref{eq:scaling-dim}. Thus the $e/8$ quasiparticle is the most relevant when $l=1$ (namely Pfaffian and PH-Pfaffian states) and should dominate the tunneling process in these two cases. When $l=3,5, 7$, tunneling for both $e/4$ and $e/8$ quasiparticles are relevant. The $e/4$ tunneling operator has  a lower scaling dimension, but what type of particles dominates the tunneling process also depends on their unrenormalized tunneling amplitudes. In the following discussion, we would assume that the tunneling process is dominated by $e/8$ quasiparticles when $l<7$. For $l>7$, the $e/8$ quasiparticle tunneling becomes irrelevant. Hence, the process is taken over by $e/4$ quasiparticles.

\subsection{$e/8$ quasiparticle tunneling} 
\label{sec:nA-e8-tunneling}

\begin{figure}[htb]
\begin{center}
\includegraphics[width=3.2 in]{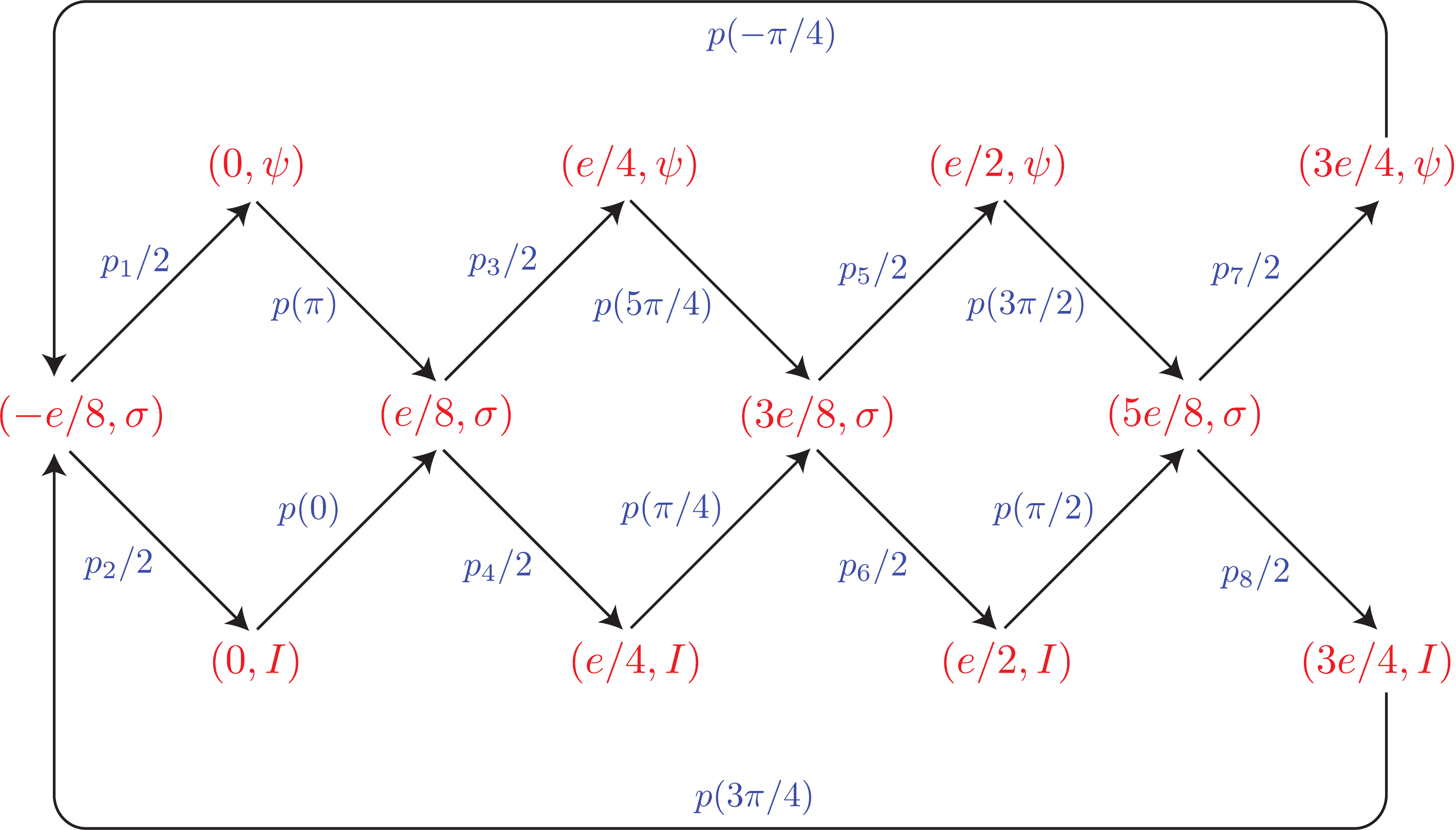}
\caption{Twelve possible superselection sectors for drain D2 when the tunneling process is dominated by charge-$e/8$ quasiparticles. The arrows show all possible transitions between different sectors at zero temperature. The corresponding transition rates and statistical phases are shown in blue. The phases in $p_i$ are listed in Table~\ref{tab:phases}.}
\label{fig:sector}
\end{center}
\end{figure}

First, we examine the tunneling current and Fano factor when the tunneling process is dominated by $e/8$ quasiparticles. In this case, there are twelve possible superselection sectors for D2, as shown in Fig.~\ref{fig:sector}. Eight consecutive tunneling events are required for the drain to absorb one electron charge. The bare transition rate between two sectors is given in Eq.~\eqref{eq:transition-rate}. Notice that some transition rates in Fig.~\ref{fig:sector} are multiplied by an additional factor of $1/2$. This additional factor comes from the fusion probability of anyons. In our case, one has equal probability of getting $\psi$ and $I$ when two vortices $\sigma$ are fused together. This probability can be calculated systematically from the algebraic theory of anyons~\cite{Kitaev}.

When an $e/8$ quasiparticle moves around an area with magnetic flux $\Phi$, an Aharanov-Bohm phase $\phi_{\rm AB}=\pi\Phi/(4\Phi_0)$ is accumulated. The symbol $\Phi_0=h/e$ denotes the magnetic flux quantum. Furthermore, the statistical phase accumulated when the quasiparticle encircles the drain D2 in the state $(ne/8, \alpha)$ is given by
\begin{eqnarray}
\phi_s
=\frac{n\pi}{8}+\phi^{\sigma\alpha}_\beta.
\end{eqnarray}
The first term comes from the U(1) bosonic charged sector, whereas the second term is contributed from the braiding between the neutral modes. The eight Chern-number-dependent phases in the transition rates between superselection sectors (see Fig.~\ref{fig:sector}) are listed in Table.~\ref{tab:phases}.

\begin{center}
\begin{table} [htb]
\begin{tabular}{| c | c || c | c |}
\hline
~~rate~~ & ~$\phi_s$~ & ~~rate~~ & ~$\phi_s$~  \\
\hline\hline
~$p_1$~ & ~$\pi\left(-1+6\mathcal{C}\right)/8$~ & ~$p_5$~ & ~$\pi\left(3+6\mathcal{C}\right)/8$~ 
\\ \hline
~$p_2$~ & ~$\pi\left(-1-2\mathcal{C}\right)/8$~ & ~$p_6$~ & ~$\pi\left(3-2\mathcal{C}\right)/8$~ 
\\ \hline 
~$p_3$~ & ~$\pi\left(1+6\mathcal{C}\right)/8$~ & ~$p_7$~ & ~$\pi\left(5+6\mathcal{C}\right)/8$~
\\ \hline
~$p_4$~ & ~$\pi\left(1-2\mathcal{C}\right)/8$~ & ~$p_8$~ & ~$\pi\left(5-2\mathcal{C}\right)/8$~ 
\\ \hline
\end{tabular}
\caption{Eight Chern-number-dependent statistical phases in the transition rates (see Fig.~\ref{fig:sector}) for the Mach-Zehnder interferometry experiment on $\nu=1/4$ FQHE. Here, the tunneling process is dominated by the $e/8$ quasiparticles.}
\label{tab:phases}
\end{table}
\end{center}

In this work, we only focus on the zero-temperature limit. Hence, quasiparticles can tunnel from the edge with the higher electrochemical potential to the edge with the lower electrochemical potential (edge 1 to edge 2 in Fig.~\ref{fig:MZI}) only. The corresponding transitions between different superselection sectors occur in one direction, as shown by the arrows in Fig.~\ref{fig:sector}. To determine the tunneling current and Fano factor for each non-Abelian order, we will employ the kinetic equation approach in Refs.~\cite{KT_noise, 16-fold}. 

To start, we introduce the symbol $P_{s,i}(t)$ for the probability that the charge $sq$ was transferred from S1 to D2 during the time $t$. Here, $q$ is the charge of the quasiparticle which dominates the tunneling process. The index $i$ labels the topological charge of drain D2 at the time $t$. The topological charge is not affected by the transfer of an integer number of electrons to D2 ($s\rightarrow s+ne/q$). The probability satisfies the following kinetic equation:
\begin{eqnarray} \label{eq:kinetic_eqn_P}
\frac{d}{dt} P_{l,i} (t)
=\sum_{j=1}^{\mathcal{N}} 
\left[
P_{l-1,j}(t)w_{j\rightarrow i} - P_{l,j} (t) w_{i\rightarrow j}
\right].
\end{eqnarray}
In the present case, $\mathcal{N}=12$, which is the number of possible superselection sectors for D2. The transition rate from sector $i$ to sector $j$ is denoted as $w_{i\rightarrow j}$. 

To proceed, we introduce a generating function
\begin{eqnarray}
f_i(z,t)
=\sum_{n=-\infty}^{\infty} P_{k+ne/q,i}(t)z^{k+ne/q}.
\end{eqnarray}
Here $k$ is uniquely determined by the topological sector $i$. From Eq.~\eqref{eq:kinetic_eqn_P}, we obtain the kinetic equation for $f_i(z,t)$:
\begin{eqnarray} \label{eq:kinetic_eqn_f}
\frac{d}{dt} f_i (z,t)
=\sum_{j=1}^{\mathcal{N} }
\left[
zf_j(z,t)w_{j\rightarrow i}
- f_i (z,t) w_{i\rightarrow j}\right].
\end{eqnarray}
The above equation can be written in the matrix form: 
$\dot{\mathbf{f}} (z,t)=\mathbf{A}\cdot\mathbf{f} (z,t)$, with $\mathbf{A}$ being a $12\times 12$ matrix. By the Rohbrach theorem~\cite{Rorbach_theorem}, all eigenvalues of $\mathbf{A}$ are non-negative at $z=1$. Also, one of them is nondegenerate and zero there. As 
$t\rightarrow\infty$, this special eigenvalue dominates the solution, which we denote as 
$\lambda(z)$.

In terms of $f_i$, the average charge being transmitted during the time interval $t$ is given by
\begin{eqnarray}
\langle Q(t)\rangle
=q\left.\left(\frac{d}{dz}\sum_{i=1}^{\mathcal{N}} f_i\right)\right|_{z=1}.
\end{eqnarray}
The tunneling current is defined as the average charge transmitted per unit time:
\begin{eqnarray} \label{eq:finite_T_current}
I
=\lim_{t\rightarrow\infty} \frac{\langle Q(t)\rangle}{t}
=q\left.\lambda'(z)\right|_{z=1}.
\end{eqnarray}
\\ \noindent 
Following the procedures in Refs.~\cite{16-fold} and~\cite{KT2008}, we obtain the tunneling current for each non-Abelian topological order:
\begin{widetext}
\begin{eqnarray} \label{eq:MZI-current}
I=\frac{er}{8}\left(\left|\Gamma_1\right|^2+\left|\Gamma_2\right|^2\right)
\left[
\frac{1-2s^2+\frac{5}{4}s^4-\frac{1}{4}s^6+\frac{1}{64}s^8\sin^2{4\gamma}}
{1-\frac{s^2}{4}\left(7+c_1\right)+\frac{5s^4}{16}\left(3+c_1\right)
-\frac{s^6}{32}\left(5+3c_1\right)
-\frac{s^8}{64}c_2(\gamma) \sin{4\gamma}}
\right].
\end{eqnarray}
\end{widetext}
Here, we have defined $\gamma=\phi_{\rm AB}+\delta$ and the parameter $s$:
\begin{eqnarray}
s=\frac{2\left|u \Gamma_1 \Gamma_2\right|} 
{\left|\Gamma_1\right|^2+\left|\Gamma_2\right|^2}.
\end{eqnarray}
Note that the condition $0\leq s \leq 1$ must be satisfied~\cite{16-fold}, so that the electric current flows from the edge with higher electrochemical potential to the lower, irrespective of 
$\phi_{\rm AB}$. Generally, there are multiple relevant operators for quasiparticle tunneling at the QPCs. All these processes contribute to the Hamiltonian in Eq.~\eqref{eq:H_tun} and affect the possible values of $s$. The renormalization group argument in Ref.~\cite{16-fold} provides a possible mechanism for achieving $s=1$ in the limit of $V\rightarrow 0$ and 
$T\rightarrow 0$ . In order to achieve the limit $s=1$, it requires $\left|\Gamma_1\right|=\left|\Gamma_2\right|$, namely a symmetric interferometer.

\begin{figure} [htb]
\includegraphics[width=3.2in]{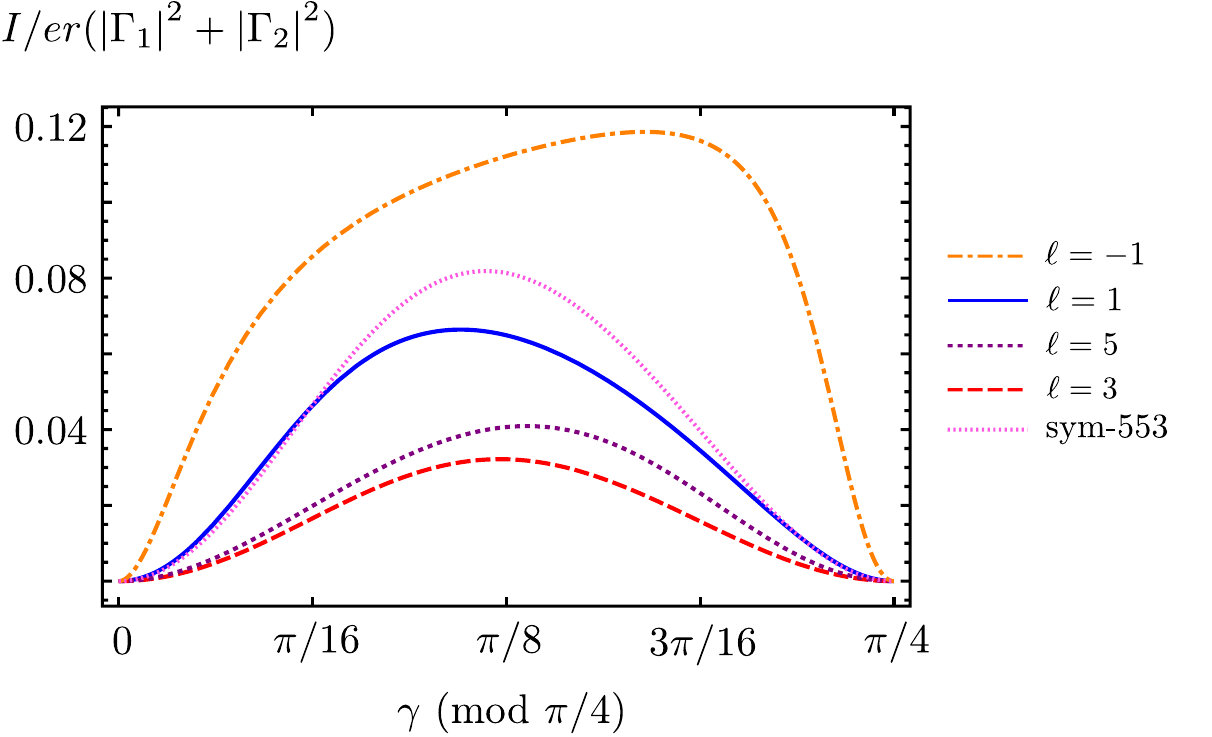}
\caption{Prediction on tunneling current in a Mach-Zehnder interferometer for different non-Abelian orders to the $\nu=1/4$ FQH state. For comparison, the result for Abelian Halperin-(5,5,3) order with flavor symmetry is also included [see Eq.~\eqref{eq:current-553-symmetric} in Sec.~\ref{sec:MZI-553}]. Here, we set $s=1$ for demonstration.}
\label{fig:MZI-current}
\end{figure}

The coefficient $c_1$ and function $c_2(\gamma)$ in Eq.~\eqref{eq:MZI-current} depend on the Chern number of the topological order (or, equivalently, pairing channel for the composite fermions). They are listed in Table~\ref{tab:c12}. In Fig.~\ref{fig:MZI-current}, we plot the tunneling current for each non-Abelian order.

\begin{center}
\begin{table} [htb]
\begin{tabular}{| c | c | c | }
\hline
~$\mathcal{C}~(\text{mod}~8)$~ & ~$c_1$~ & ~$c_2(\gamma)$~  \\
\hline\hline
~1~ & ~$\sin{(\pi/8)}$~ & ~$cos{(13\pi/16)}\sin{(4\gamma-3\pi/16)}$~ \\ \hline
~-1~ & ~$\cos{(\pi/8)}$~ & ~$\cos{(15\pi/16)}\sin{(4\gamma+\pi/16)}$~ \\ \hline 
~3~ & ~$-\cos{(\pi/8)}$~ & ~$\sin{(\pi/16)}\cos{(4\gamma+\pi/16)}$~ \\ \hline
~5~ & ~$-\sin{(\pi/8)}$~  & ~$\sin{(19\pi/16)}\cos{(4\gamma-3\pi/16)}$~ \\ \hline
\end{tabular}
\caption{The coefficient $c_1$ and the function $c_2(\gamma)$ in Eq.~\eqref{eq:MZI-current} for different Chern numbers, $\mathcal{C}$. It is reminded that $\mathcal{C}=\ell$ as proven in Appendix~\ref{app:Chern}.}
\label{tab:c12}
\end{table}
\end{center}

The corresponding Fano noise to the tunneling current is defined as the following autocorrelation function:
\begin{eqnarray}
S(\omega)
=\frac{1}{2}\int_{-\infty}^\infty
\big\langle
I(0)I(t)+I(t)I(0)\big\rangle e^{i\omega t}~ dt.
\end{eqnarray}
Our definition follows the convention in Ref.~\cite{KT_noise}, such that a prefactor $1/2$ is included. In the low-frequency limit, the Fano noise and the tunneling current are not independent. In fact, they satisfy the following relation:
\begin{eqnarray}
e^*=S/I.
\end{eqnarray}
The ratio $e^*$ is known as the Fano factor. It can be evaluated as~\cite{16-fold, KT2008}
\begin{eqnarray} \label{eq:finite_T_Fano}
e^*
=\lim_{t\rightarrow\infty} 
\frac{\langle \delta Q^2(t)\rangle}{\langle Q(t)\rangle}
=q\left[1+\frac{\left.\lambda''(z)\right|_{z=1}}
{\left.\lambda'(z)\right|_{z=1}}\right].
\end{eqnarray}
Here, $\langle \delta Q^2(t)\rangle$ is the variance of the average charge transmitted in the time interval $t$. It can be obtained from $f_i$ as follows:
\begin{eqnarray}
\langle \delta Q^2(t)\rangle
=q^2\left.\left(\frac{d}{dz}z\frac{d}{dz}\sum_{i=1}^{\mathcal{N}} f_i\right)\right|_{z=1}
-\langle Q(t)\rangle^2 .
\end{eqnarray}

\begin{figure} [htb]
\includegraphics[width=3.2in]{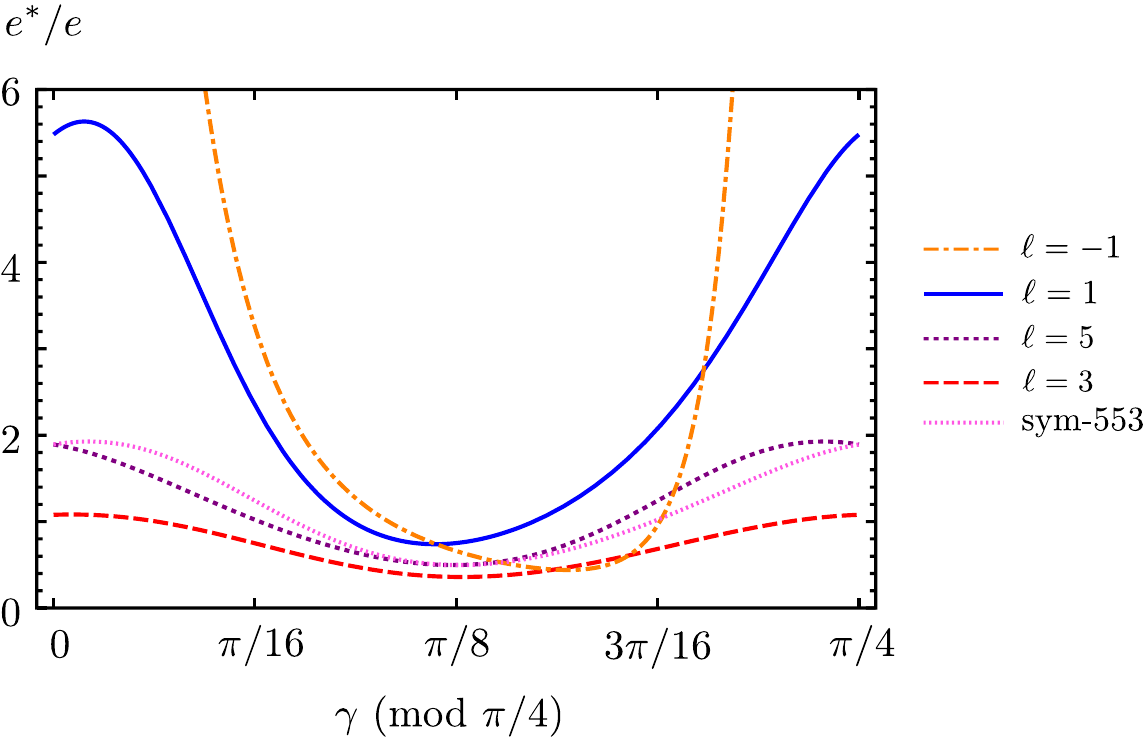}
\caption{Prediction on Fano factor in a Mach-Zehnder interferometer for different non-Abelian orders to the $\nu=1/4$ FQH state. The result for Abelian Halperin-(5,5,3) order with flavor symmetry is also included for comparison. Here, we set $s=1$ for demonstration.}
\label{fig:MZI-noise}
\end{figure}

Similar to the tunneling current, it is straightforward to evaluate the Fano factor for each non-Abelian topological order. However, the general expression is too lengthy to be displayed here. The maximum Fano factor is achieved at $s=1$. In Fig.~\ref{fig:MZI-noise}, we set $s=1$ and plot the Fano factor against $\gamma$ for each non-Abelian topological order. It is observed that both tunneling current and Fano factor are periodic in $\gamma$ with a period of $\pi/4$. This feature is consistent with the Byers-Yang theorem~\cite{Byers-Yang}. In addition, the maximum Fano factor at $s=1$ for each topological order has been determined numerically. The results are listed in Table~\ref{tab:max_Fano}.

\begin{center} 
\begin{table} [htb]
\begin{tabular} {| c || c | c || c | c |}
\hline
~$\ell$~ &
~$(e^*/e)_{\rm max}$~ & ~$\gamma_{\rm max}$~ 
& ~$(e^*/e)_{\rm min}$~ & ~$\gamma_{\rm min}$~  \\
\hline\hline
~$1$~ & ~$5.63$~ & ~$0.03$~ & $0.74$ & ~$0.37$~ \\ \hline
~$-1$~ &~$53.2$~ & ~$0.77$~ & ~$0.44$~ & ~$0.50$~ \\ \hline
~$3$~ &~$1.08$~ & ~$0.02$~ & ~$0.36$~ & ~$0.40$~ \\ \hline 
~$5$~ &~$1.93$~ & ~$0.75$~ & ~$0.50$~ & ~$0.39$~ \\ \hline
\end{tabular}
\caption{Extremal values for Fano factor at $s=1$. Here, $\gamma_{\rm max}$ and 
$\gamma_{\rm min}$ are the optimal values for the Fano factor to achieve the extremal values. Notice that both $\gamma_{\rm max}$ and $\gamma_{\rm min}$ are modulo $\pi/4$.}
\label{tab:max_Fano}
\end{table}
\end{center}

\subsubsection{Signatures for 22111 parton order in $\nu=1/4$ FQHE}

When the width of the quantum well and electron density of the system are sufficiently large, it was recently suggested that the 22111 parton order may describe the ground state of the $\nu=1/4$ FQHE~\cite{wide-well}. This speculated parton order is topologically equivalent to an $f$-wave paired state of composite fermions, i.e., pairing composite fermions in the $\ell=3$ channel. In this case, it is observed from Fig.~\ref{fig:MZI-current} that the tunneling current is nearly (but not truly) symmetric about $\gamma =\pi/8~(\text{mod}~\pi/4)$. This feature is absent in other paired states. In addition, the maximum Fano factor at $s=1$ is found to be about $1.08$. As shown in Fig.~\ref{fig:MZI-noise}, it is rather likely for other topological orders to exceed this maximal value. Therefore, we suggest both the tunneling current and Fano factor measurement in Mach-Zehnder interferometry can provide tight constraints to identify the parton order.

\subsubsection{Signatures for PH-Pfaffian state in $\nu=1/4$ FQHE}

Aside from the $\ell=3$ paired state, it is possible for the Mach-Zehnder interferometer to identify the PH-Pfaffian order ($\ell=-1$) for the $\nu=1/4$ FQH state. As shown in Fig.~\ref{fig:MZI-current}, the tunneling current can reach a maximum value of $0.12$ [in units of $er (\left|\Gamma_1\right|^2+\left|\Gamma_2\right|^2)$]. This value is at least $50\%$ larger than the maximum current that can be achieved by other topological orders. In addition, Fig.~\ref{fig:MZI-noise} shows that the Fano factor increases rapidly at $\gamma\approx\pi/16~(\text{mod}~\pi/8)$ when $s=1$. It reaches a maximum value of about $53.2$. This extremal value is one order of magnitude larger than the corresponding values for other paired states.

\subsection{$e/4$ quasiparticle tunneling}

\begin{figure}[htb]
\begin{center}
\includegraphics[width=3.3 in]{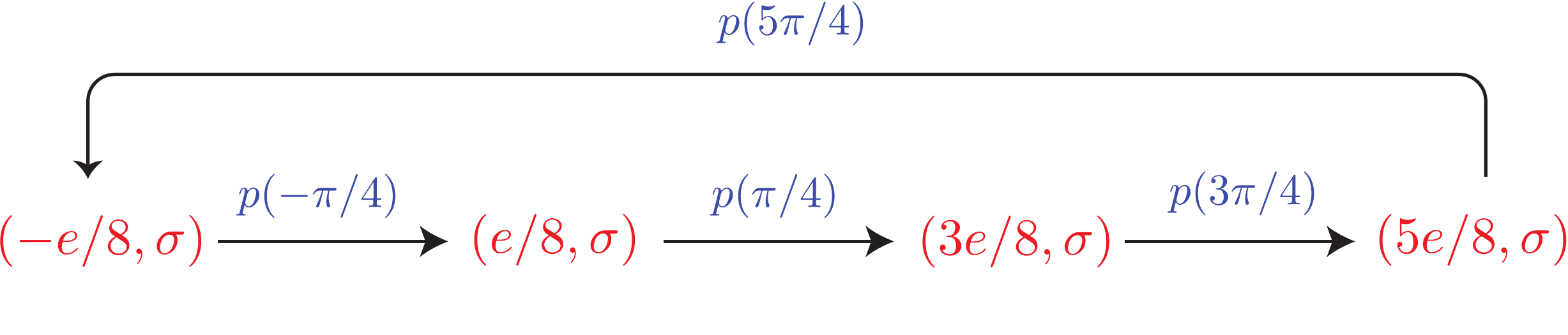}
\caption{Four possible superselection sectors for drain D2 when the tunneling is dominated by charge-$e/4$ quasiparticles with topological charge $I$. Here, D2 has an odd number of $e/8$ quasiparticles. The arrows show all possible transitions between different sectors at zero temperature. The corresponding transition rates and statistical phases are shown in blue.}
\label{fig:sector-e4-I}
\end{center}
\end{figure}

\begin{figure}[htb]
\begin{center}
\includegraphics[width=3.3 in]{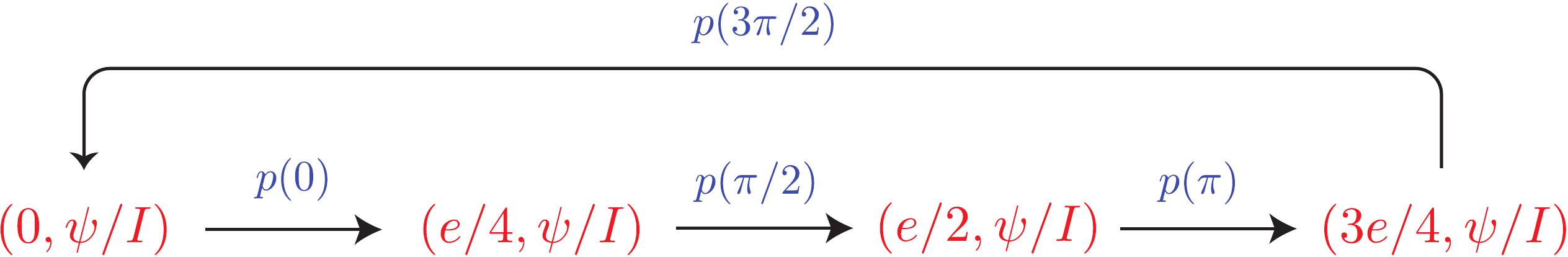}
\caption{Four possible superselection sectors for drain D2 when the tunneling is dominated by charge-$e/4$ quasiparticles with topological charge $I$. Here, D2 has an even number of $e/8$ quasiparticles. The arrows show all possible transitions between different sectors at zero temperature. The corresponding transition rates and statistical phases are shown in blue.}
\label{fig:sector-e4-II}
\end{center}
\end{figure}

We complete our analysis on Mach-Zehnder interferometry for non-Abelian orders with a short discussion on $e/4$ quasiparticle tunneling. Depending on the number of $e/8$ quasiparticles in the drain D2, two different scenarios may occur. We illustrate these two cases separately in Figs.~\ref{fig:sector-e4-I} and~\ref{fig:sector-e4-II}. The most relevant charge-$e/4$ quasiparticles have a trivial topological charge. The quasiparticle is described by the vertex operator $\Psi_{e/4}=e^{i\varphi_\rho}$. When it encircles the drain D2, a statistical phase
\begin{eqnarray}
\phi_s'
=\frac{n\pi}{4}
\end{eqnarray} 
is accumulated. The symbol $n$ denotes the number of charge-$e/8$ quasiparticles in D2. It is important to notice that $\phi_s'$ does not depend on the pairing channel of the composite fermions.

Following similar procedures in previous discussion, one can set up a new set of kinetic equations to determine the tunneling current and Fano factor. When D2 has an odd number of $e/8$ quasiparticles, the tunneling current takes the form
\begin{eqnarray}
I=\frac{er}{4}\left(\left|\Gamma_1\right|^2 + \left|\Gamma_2\right|^2\right)
\left[\frac{1-s^2+\frac{s^4}{8}\left(1+\cos{4\gamma'}\right)}{1-\frac{s^2}{2}}\right].
\end{eqnarray}
Meanwhile, the Fano factor is given by
\begin{eqnarray}
\frac{e^*}{e}
=\frac{1-\frac{s^2}{2}+\frac{s^4}{8}+\frac{s^6}{16}+\frac{s^4}{16}\left(s^2-6\right)\cos{4\gamma'}}
{4\left(1-\frac{s^2}{2}\right)^2}.
\end{eqnarray}
Here, all symbols $\Gamma_1$, $\Gamma_2$, $s$ are defined for the tunneling process of $e/4$ quasiparticles. Also, we define $\gamma'=\pi\Phi/(2\Phi_0)+\delta$. 

On the other hand, the tunneling current and Fano factor when $n$ is even are given by
\begin{eqnarray}
I=\frac{er}{4}\left(\left|\Gamma_1\right|^2 + \left|\Gamma_2\right|^2\right)
\left[\frac{1-s^2+\frac{s^4}{8}\left(1-\cos{4\gamma'}\right)}{1-\frac{s^2}{2}}\right]
\end{eqnarray}
and
\begin{eqnarray}
\frac{e^*}{e}
=\frac{1-\frac{s^2}{2}+\frac{s^4}{8}+\frac{s^6}{16}-\frac{s^4}{16}\left(s^2-6\right)\cos{4\gamma'}}
{4\left(1-\frac{s^2}{2}\right)^2}.
\end{eqnarray}
Notice that the expressions for the two cases are shifted by a phase of $\pi/4$ due to an additional 
$e/8$ quasiparticle in D2. Also, the period of $\pi/2$ in both tunneling current and Fano factor are expected~\cite{Byers-Yang}. Since the four sectors in D2 are connected as in the Laughlin states, the maximum Fano factor is $e^*=e$ at $s=1$~\cite{KT_noise}. 

In a general situation, all $e/4$, $e/8$ quasiparticles and the neutral fermion can tunnel at the QPCs. The corresponding tunneling current and Fano factor can be determined by solving a full set of kinetic equations. This procedure is straightforward but beyond the scope of our current manuscript.

\section{Experimental signatures of Halperin-(5,5,3) order}
\label{sec:553}

Although we focused mainly on non-Abelian topological orders in the previous two sections, the possibility of having a two-component Abelian order in the $\nu=1/4$ FQH state has not been completely ruled out. On the contrary, two-component orders become favorable if the two effective layers of 2DEG in a wide quantum well have a weak interlayer tunneling. Thus, it is equally important to examine the experiment signatures for two-component orders for the 
$\nu=1/4$ FQH state.

In this section, we concentrate on the spin-unpolarized Halperin-(5,5,3) order. It was suggested that this Abelian order is a strong competitor to the non-Abelian orders in describing the 
$\nu=1/4$ FQH state in a wide quantum well~\cite{Papic-1/4}. Other two-component candidates, such as Halperin-(7,7,1) and Halperin-(5,13,1) orders are rather unlikely to be the solution. In particular, the former assumes each layer of electron gas has a filling factor of $1/7$. With such a low filling factor, it is likely for the 2D electron gas to host a coupled Wigner crystal rather than a FQH state~\cite{exp-wide-well1}. For the latter, it requires a strong density imbalance in the two effective layers. Furthermore, the Halperin-(5,13,1) order was also eliminated by the numerical results in Ref.~\cite{Papic-1/4} due to its requirement of having an unrealistically wide quantum well. In Appendix~\ref{app:NA-2C}, a brief discussion of several other two-component candidates is provided. However, the numerical results in Ref.~\cite{wide-well} suggest that they are unlikely to be realized in the GaAs quantum well setup. 

\subsection{Edge structure and thermal Hall conductance}

We start our discussion by reviewing the edge physics of the Halperin-(5,5,3) order, which is described by the following Lagrangian density:
\begin{eqnarray}
\mathcal{L}
=-\frac{1}{4\pi}\sum_{i,j}
\left[
K_{ij}\partial_t\varphi_i \partial_x \varphi_j
+V_{ij}\partial_x\varphi_i\partial_x\varphi_j\right].
\end{eqnarray}
The corresponding two-by-two $K$ matrix and charge vector $t$ are
\begin{eqnarray}  \label{eq:K-553}
K=
\begin{pmatrix}
5 & 3 \\
3 & 5\\
\end{pmatrix}
~,~
t=
\begin{pmatrix}
1  \\ 1
\end{pmatrix}.
\end{eqnarray}
The $V$ matrix characterizes the interaction between the two edge modes. The edge of the topological order has two downstream bosonic charged modes, $\varphi_1$ and $\varphi_2$. Thus the thermal Hall conductance is predicted to be
\begin{eqnarray}
\kappa_H
=2\left(\frac{\pi^2 k_B^2 T}{3h}\right).
\end{eqnarray}

\subsection{Quasiparticles and tunneling exponents}

Generically, any quasiparticle in an Abelian two-component topological order can be represented by a vertex operator~\cite{Wen_book}:
\begin{eqnarray} 
\Psi_q
=e^{i\left(l_1\varphi_1 + l_2\varphi_2\right)}
=e^{i\bm{l}\cdot\bm{\varphi}}.
\end{eqnarray}
Here, we define $\bm{l}=(l_1,l_2)\in\mathbb{Z}^2$ and $\bm{\varphi}=(\varphi_1, \varphi_2)$. The quasiparticle has charge $q$:
\begin{eqnarray}
q=e(\bm{l}^T K^{-1} t).
\end{eqnarray}
Since all edge modes propagate in the same direction, the scaling dimension of $\Psi_q$ is independent of the interaction between the edge modes. Specifically, one has 
\begin{eqnarray}
\Delta_q=\frac{1}{2}\left(\bm{l}^T K^{-1}\bm{l}\right).
\end{eqnarray}
Furthermore, a phase of $\phi_{12}=2\pi \bm{l}_1^T K^{-1}\bm{l}_2$ is accumulated when a quasiparticle characterized by $\bm{l}_1$ encircles another quasiparticle characterized by 
$\bm{l}_2$, in the counterclockwise direction. 

The two most relevant electron operators for the spin-unpolarized Halperin-(5,5,3) order are given by
\begin{eqnarray} \label{eq:553-electron}
\Psi_e
=e^{5i\varphi_1+ 3i\varphi_2}
\quad\text{and}\quad
\Psi_e
=e^{3i\varphi_1+ 5i\varphi_2}.
\end{eqnarray}
Both of them have scaling dimension $\Delta_e=5/2$. Different from non-Abelian orders, there are two types of the most fundamental quasiparticles. They are described by the vertex operators
\begin{eqnarray} \label{eq:op-553}
\Psi_{e/8}
=e^{i\varphi_1}
\quad\text{and}\quad
\Psi_{e/8}
=e^{i\varphi_2}.
\end{eqnarray}
Both of them have charge $e/8$ and scaling dimension $\Delta_{e/8}=5/32$. For convenience in later discussion, we simply name the $e/8$ quasiparticle described by $\mathbf{a}=(1,0)$ and $\mathbf{b}=(0,1)$ as $\mathbf{a}$ and $\mathbf{b}$ quasiparticles, respectively. It is straightforward to verify that the two operators in Eq.~\eqref{eq:op-553} have single-valued OPEs with the two electron operators in Eq.~\eqref{eq:553-electron}. Lastly, we remind that the charge-$e/4$ qausiparticles are characterized by the vector $\bm{l}=(1,1)$. Equivalently, they are described by the vertex operator:
\begin{eqnarray}
\Psi_{e/4}
=e^{i\varphi_1}e^{i\varphi_2},
\end{eqnarray}
which has scaling dimension $\Delta_{e/4}=1/8$.

\subsubsection{Fractional statistics}

Now, we determine the phase accumulated when an $e/8$ quasiparticle encircles another 
$e/8$ quasiparticle. When the two quasiparticles are identical, one has
\begin{eqnarray}
\phi_{11}=\phi_{22}
=\frac{5\pi}{8}.
\end{eqnarray}
If the two quasiparticles are different, then the mutual statistical phase is
\begin{eqnarray}
\phi_{12}
=-\frac{3\pi}{8}.
\end{eqnarray}
These two results are important to our discussion on Mach-Zehnder interferometry in the next subsection.

\subsubsection{Tunneling exponents}

From the previous discussion on scaling dimensions, the tunneling exponents for $e/8$, $e/4$ quasiparticles and electron for the Halperin-(5,5,3) order are
\begin{eqnarray}
g_{e/8}=\frac{5}{16}
~,~
g_{e/4}=\frac{1}{4}
~,~
g_e=5.
\end{eqnarray}
Notice that $g_{e/8}$ are different from all tunneling exponents predicted for one-component non-Abelian orders in Sec.~\ref{sec:wavefunction-NA} (see Table~\ref{tab:summary} also). Thus the tunneling experiment may distinguish between the Halperin-(5,5,3) order and other non-Abelian orders, given that the tunneling process is dominated by $e/8$ quasiparticles.

\subsection{Mach-Zehnder interferometry}
\label{sec:MZI-553}

Following the renormalization group analysis in Sec.~\ref{sec:RG-analysis}, both charge-$e/8$ and charge-$e/4$ quasiparticle tunneling are relevant processes for the Halperin-(5,5,3) order. Again, we will assume the process is dominated by the $e/8$ quasiparticles in the following discussion. For the present case, we need to take care of the two flavors of $e/8$ quasiparticles, namely the $\mathbf{a}=(1,0)$ and $\mathbf{b}=(0,1)$ quasiparticles. In a general situation, they have different tunneling amplitudes at the quantum point contacts. Also, the probability of exciting them in the FQH system can be different. A special case arises if an exact or approximate flavor symmetry exists between the $\mathbf{a}$ and $\mathbf{b}$ quasiparticles. Then, the Abelian topological order can also demonstrate the even-odd effect in a Fabry-P\'{e}rot interferometer, with the same reasoning in the case of Halperin-(3,3,1) order~\cite{Stern_PRB2010}. In other words, the observation of even-odd effect is not a decisive experimental signature for identifying a one-component non-Abelian order. This subtlety motivates us to examine Mach-Zehnder interferometry on the (5,5,3) order. Our analysis follows closely to previous work on the (3,3,1) order~\cite{Chenjie2010} and (1,1,3) order~\cite{Guang2015, foot-335}. At the end, we find that both predicted tunneling current and Fano factor for the (5,5,3) order are different from those results for non-Abelian orders in Sec.~\ref{sec:MZ-exp}.

\begin{figure}[htb]
\begin{center}
\includegraphics[width=3.3 in]{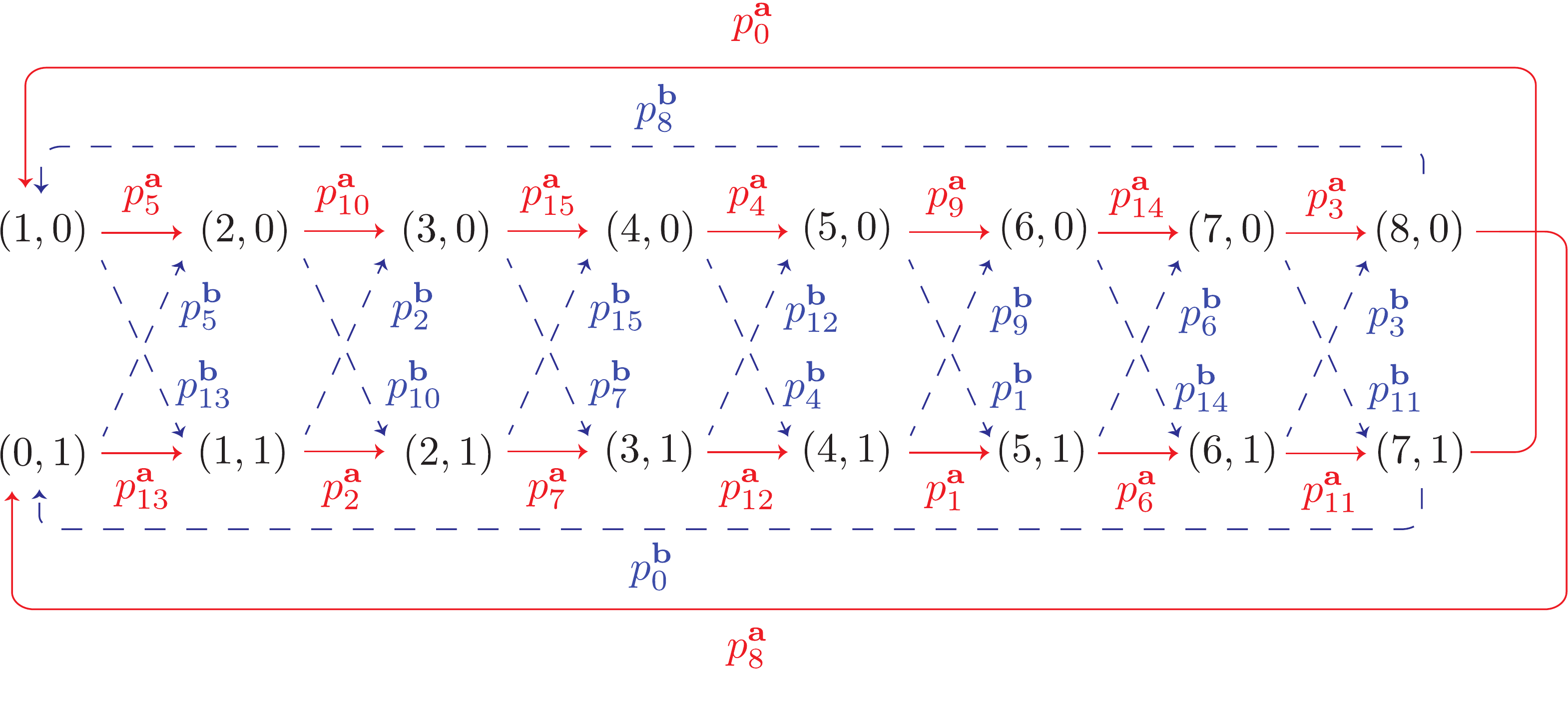}
\caption{Sixteen possible superselection sectors for drain D2 in the Mach-Zehnder interferometer when the tunneling process is dominated by $e/8$ quasiparticles in the Halperin-(5,5,3) order. The red solid lines and blue dashed lines show the transitions between different sectors when the incoming $e/8$ quasiparticle is an $\mathbf{a}=(1,0)$ and a $\mathbf{b}=(0,1)$ quasiparticle, respectively. The corresponding transition rates are either $p^{\mathbf{a}}_j$ or 
$p^{\mathbf{b}}_j$, as defined in Eq.~\eqref{eq:rate-553}.}
\label{fig:sector-553}
\end{center}
\end{figure}

Consider the situation when there are $m$ copies of $\mathbf{a}$ and $n$ copies of $\mathbf{b}$ quasiparticles being stored in the drain D2 (see Fig.~\ref{fig:MZI} for the experimental setup). We denote these superselection sectors by a vector $\bm{l}=(m,n)$. The electric charge in D2 is 
$e(m+n)/8$. When an incoming $\mathbf{a}$ quasiparticle encircles the drain D2, a phase of
\begin{eqnarray}
\phi_s^{\bm{a}}
=\frac{5\pi m}{8}-\frac{3\pi n}{8}
=\frac{\pi}{8}\left(5m-3n\right)
\end{eqnarray}
will be accumulated. If the incoming particle is a $\mathbf{b}$ quasiparticle, then the corresponding phase becomes
\begin{eqnarray}
\phi_s^{\bm{b}}
=\frac{5\pi n}{8}-\frac{3\pi m}{8}
=\frac{\pi}{8}\left(5n-3m\right).
\end{eqnarray}
Furthermore, two particles with $\bm{l}$ and $\bm{l}'=\bm{l}+n_1(5,3)+n_2(3,5)$ are identified. It is because the same phase would be accumulated when an $e/8$ quasiparticle encircles them. As a result, D2 can have 16 different possible superselection sectors as illustrated in Fig.~\ref{fig:sector-553}. Importantly, all 16 sectors are connected. Otherwise, processes described by less relevant operators matter. 

Depending on the flavors of the incoming $e/8$ quasiparticle, $\mathbf{x}=\mathbf{a}$ or $\mathbf{b}$,  the transition rates between the superselection sectors are
\begin{align} \label{eq:rate-553}
\nonumber
p^{\mathbf{x}}_j
=r\left(\left|\Gamma_1^{\mathbf{x}}\right|^2 + \left|\Gamma_2^{\mathbf{x}}\right|^2\right)
\left[
1+s^{\mathbf{x}}\cos{\left(\frac{\pi\Phi}{4\Phi_0} +\frac{j\pi}{8}+\delta^{\mathbf{x}}\right)}\right].
\\
\end{align}
Here, the two symbols
$s^{\mathbf{x}}=2\left|u\Gamma_1^{\mathbf{x}} \Gamma_2^{\mathbf{x}}\right| /
\left(\left|\Gamma_1^{\mathbf{x}}\right|^2+\left|\Gamma_2^{\mathbf{x}}\right|^2\right)$ and 
$\delta^{\mathbf{x}}=\text{arg}\left(u\Gamma_2^{\mathbf{x}}/\Gamma_1^{\mathbf{x}}\right)$ are defined.

\begin{figure}[htb]
\begin{center}
\includegraphics[width=3.3 in]{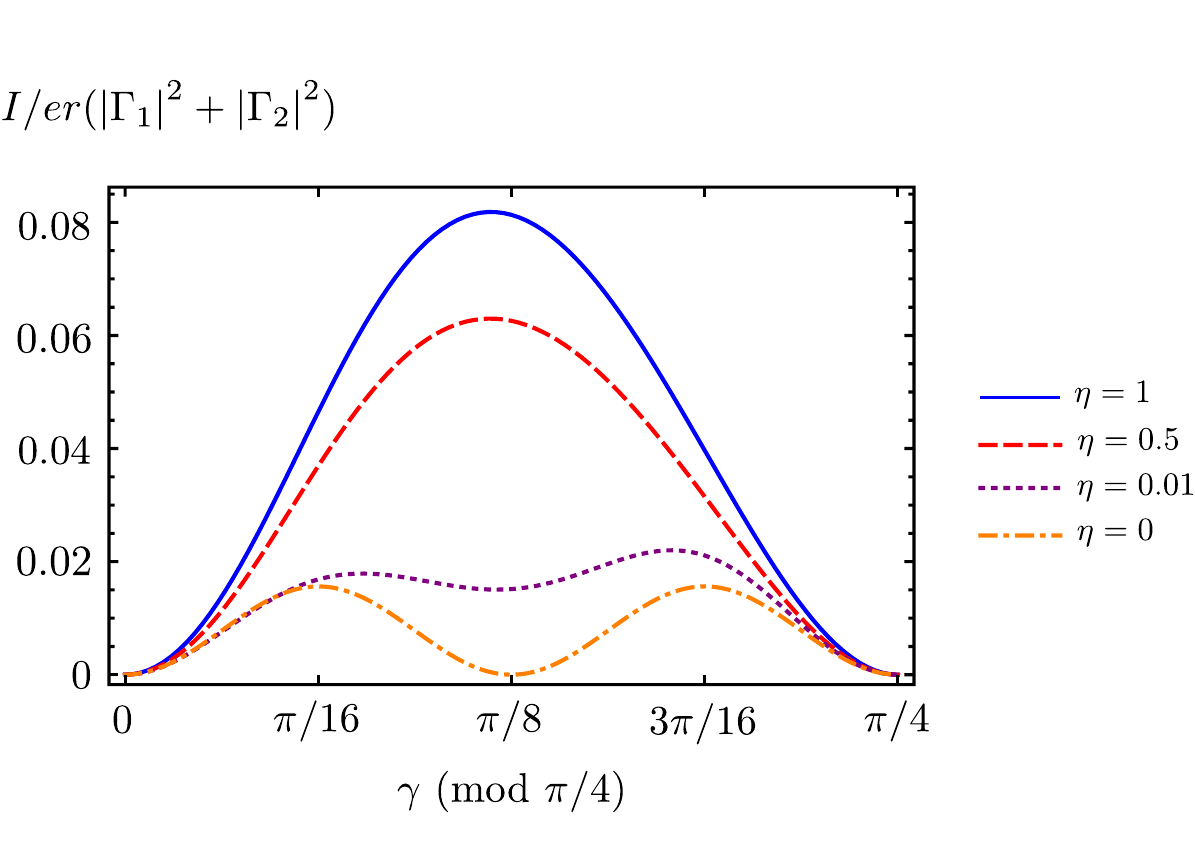} 
\caption{Prediction on tunneling current in a Mach-Zehnder interferometer for Halperin-(5,5,3) order. In the plot, we have set $s^{\mathbf{a}}=s^{\mathbf{b}}=1$ and $\delta^{\mathbf{a}}=\delta^{\mathbf{b}}$ in Eq.~\eqref{eq:rate-553}. Different curves correspond to different values of 
$\eta=\left|\Gamma^{\mathbf{b}}_i\right|^2/\left|\Gamma^{\mathbf{a}}_i\right|^2$.}
\label{fig:current-553}
\end{center}
\end{figure}

\begin{figure}[htb]
\begin{center}
\includegraphics[width=3.3 in]{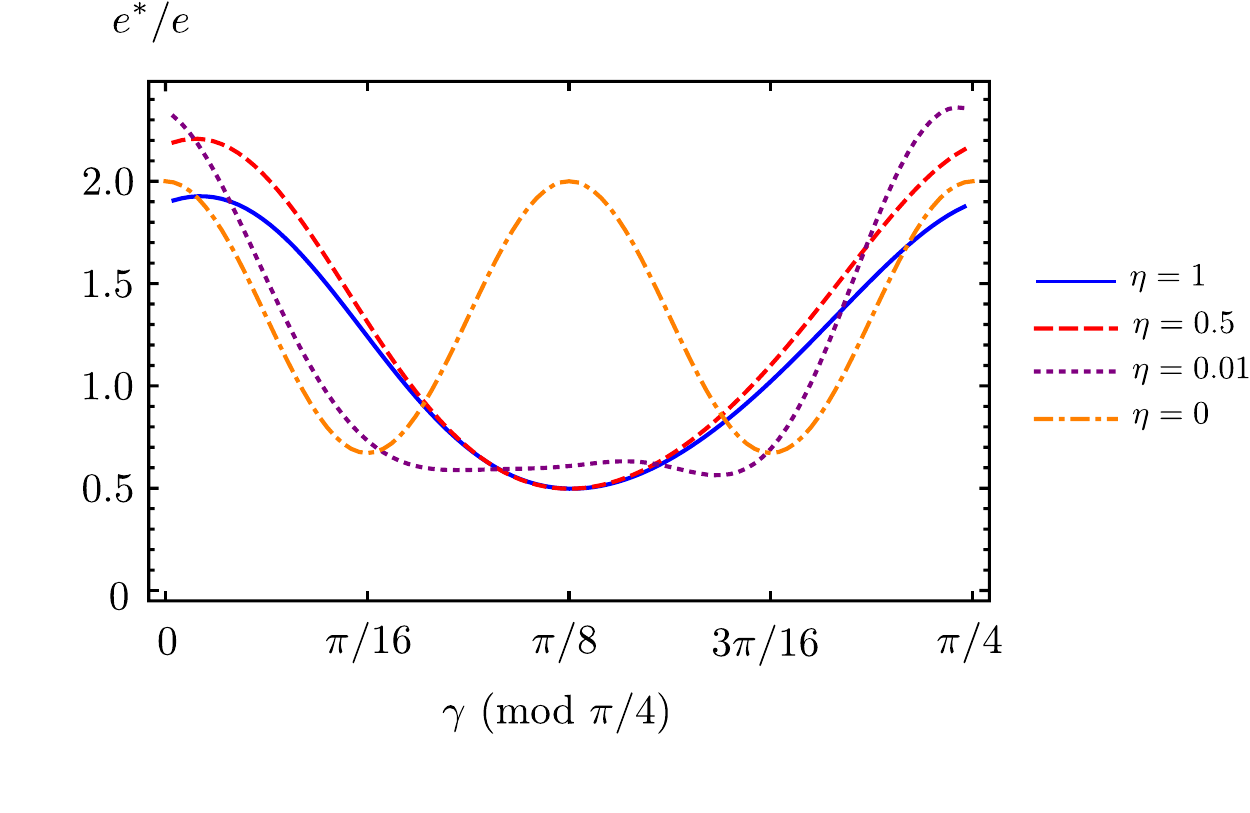} 
\caption{Prediction on Fano factor in a Mach-Zehnder interferometer for Halperin-(5,5,3) order. Here, we set $s^{\mathbf{a}}=s^{\mathbf{b}}=1$ and $\delta^{\mathbf{a}}=\delta^{\mathbf{b}}$ in Eq.~\eqref{eq:rate-553}. Different curves correspond to different values of 
$\eta=\left|\Gamma^{\mathbf{b}}_i\right|^2/\left|\Gamma^{\mathbf{a}}_i\right|^2$.}
\label{fig:Fano-553}
\end{center}
\end{figure}

The tunneling current and Fano factor can be obtained from the kinetic equation approach in Sec.~\ref{sec:nA-e8-tunneling} by formulating a new $16\times 16$ matrix $\mathbf{A}$ to describe the transition rates between the superselection sectors shown in Fig.~\ref{fig:sector-553}. However, the general expressions are very lengthy to display here. In order to simplify our discussion and highlight some special cases, we set $s^{\mathbf{a}}=s^{\mathbf{b}}=1$ and $\delta^{\mathbf{a}}=\delta^{\mathbf{b}}$. Also, we define $\left|\Gamma_i^{\mathbf{b}}\right|^2=\eta
\left|\Gamma_i^{\mathbf{a}}\right|^2$, where $i=1,2$ labels the QPCs. Here, the parameter $\eta$ characterizes the asymmetry between the two flavors of $e/8$ quasiparticles in the topological order. We show the tunneling current and Fano factor for several values of $\eta$ in Fig.~\ref{fig:current-553} and Fig.~\ref{fig:Fano-553}, respectively. 

\subsubsection{Quasiparticles with flavor symmetry}

\begin{figure}[htb]
\begin{center}
\includegraphics[width=3.3 in]{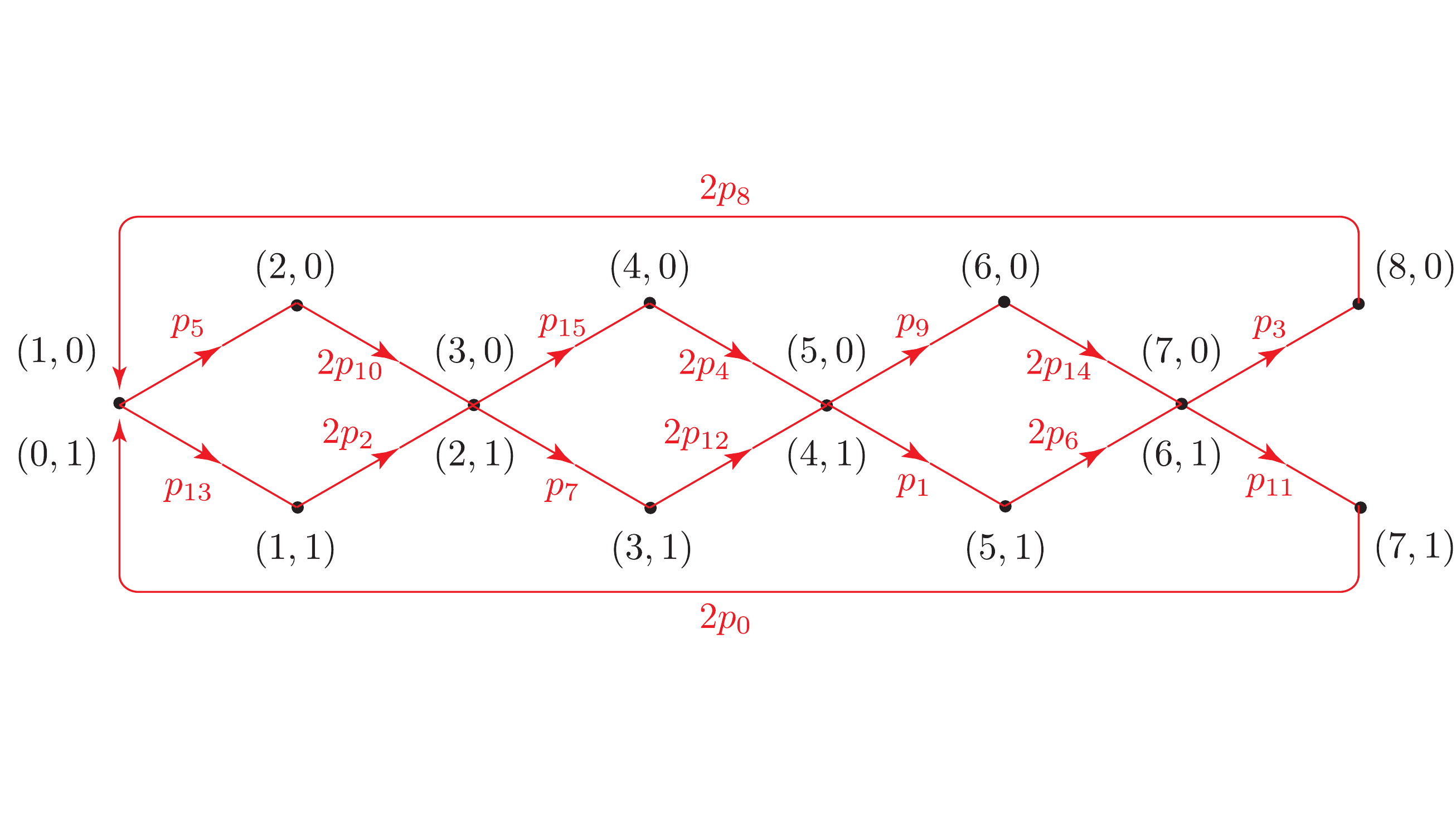}
\caption{Superselection sectors for D2 when an exact flavor symmetry exists for the 
$\mathbf{a}=(1,0)$ and $\mathbf{b}=(0,1)$ quasiparticles in the Halperin-(5,5,3) order. The transition rates satisfy $p^{\mathbf{a}}_j=p^{\mathbf{b}}_j\equiv p_j$.}
\label{fig:sector-553-symmetry}
\end{center}
\end{figure}

Suppose there is an exact flavor symmetry between the $\mathbf{a}$ and $\mathbf{b}$ types of $e/8$ quasiparticles. This scenario is captured by the setting of $\eta=1$. Then, some superselection sectors for D2 in Fig.~\ref{fig:sector-553} are identified. The end result is shown in Fig.~\ref{fig:sector-553-symmetry}. In this case, the tunneling amplitudes satisfy 
$\Gamma^{\mathbf{a}}_i=\Gamma^{\mathbf{b}}_i$. Thus the transition rates in Eq.~\eqref{eq:rate-553} simplify to  $p^{\mathbf{a}}_j=p^{\mathbf{b}}_j=p_j$. 

The corresponding tunneling current is given by Eq.~\eqref{eq:current-553-symmetric}, which has a similar form to Eq.~\eqref{eq:MZI-current}. This similarity can be understood since both Fig.~\ref{fig:sector} and Fig.~\ref{fig:sector-553} have the same topology. In order to compare with the results for non-Abelian orders, we also plot Eq.~\eqref{eq:current-553-symmetric} in Fig.~\ref{fig:MZI-current}. From the figure, it is observed that the overall shape for the tunneling current for the Halperin-(5,5,3) order is different from the results for non-Abelian orders. This feature suggests that it can be more effective to distinguish Abelian and non-Abelian orders for the 
$\nu=1/4$ FQH state by performing Mach-Zehnder experiment.

\begin{widetext}
\begin{eqnarray} \label{eq:current-553-symmetric}
I=\frac{er}{4}\left(\left|\Gamma_1\right|^2+\left|\Gamma_2\right|^2\right)
\left[
\frac{1-2s^2+\frac{5}{4}s^4-\frac{1}{4}s^6+\frac{1}{64}s^8\sin^2{4\gamma}}
{1-\frac{s^2}{4}\left(7-\sin{\frac{\pi}{8}}\right)+\frac{5s^4}{16}\left(3-\sin{\frac{\pi}{8}}\right)
-\frac{s^6}{32}\left(5-3\sin{\frac{\pi}{8}}\right)
-\frac{s^8}{64}\sin{\frac{3\pi}{16}}\cos{(\frac{3\pi}{16}+4\gamma)} \sin{4\gamma}}
\right].
\end{eqnarray}
\end{widetext}

Furthermore, the maximum and minimum Fano factors when $s=1$ are found to be 
$e^*_{\rm max}\approx 1.93e$ and $e^*_{\rm min}\approx 0.50e$. These two values are very close to the results for the $(\ell=5)$-paired state. Nevertheless, the overall shape of the two curves as a function of $\gamma$ are not identical. This is illustrated in Fig.~\ref{fig:MZI-noise}.

\subsubsection{Tunneling by only one flavor of quasiparticles}

Another special case happens when only one flavor of $e/8$ quasiparticles is allowed to tunnel at the QPCs. Suppose this quasiparticle is the $\mathbf{a}=(1,0)$ particle. Then, the scenario is captured by the setting of $\eta=0$. In this case, the 16 superselection sectors for D2 are connected in a simple way, analogous to the Laughlin state (simply connected by the red solid lines in Fig.~\ref{fig:sector-553}). In other words, it takes 16 consecutive $e/8$-quasiparticle tunneling events for the drain to return to its initial state. As a result, the periods in the tunneling current and Fano factor reduce to $\pi/8$. Also, the Fano factor has a maximum value of $e^*=2e$~\cite{16-fold}.

From Figs.~\ref{fig:MZI-current}, \ref{fig:MZI-noise}, \ref{fig:current-553}, and~\ref{fig:Fano-553}, we find that the difference between the experimental signatures in Mach-Zehnder interferometry for Halperin-(5,5,3) order and non-Abelian orders become more transparent in the limit $\eta\rightarrow 0$. On the other hand, it becomes more challenging to resolve the small difference when $\eta\rightarrow 1$. Therefore, it requires a combination of different types of experiment to identify the topological order in the $\nu=1/4$ FQH state.

\section{Summary of experimental signatures}
\label{sec:summary-exp}

In Table~\ref{tab:summary}, we summarize the experimental signatures for different topological orders. Based on the table, we comment briefly on how the results may help to identify the nature of the $\nu=1/4$ FQH state. 

First, consider the tunneling experiment. We assume the tunneling process is dominated by the smallest-charge quasiparticles-in other words, the charge-$e/8$ quasiparticles. Under this assumption, topological orders with different numbers of Majorana modes at the edge will have different tunneling exponents $g_{e/8}$. However, the chirality of the Majorana modes cannot be determined from tunneling experiment. In order to differentiate between topological orders with upstream and downstream Majorana modes, an  additional experiment is required. This complementary experiment can be upstream noise probing experiment or thermal Hall conductance measurement. If topologically protected upstream neutral modes are observed, then it provides a support to the PH-Pfaffian and anti-Pfaffian orders. A Mach-Zehnder experiment or thermal Hall conductance measurement may differentiate between these two orders. 

For topological orders having more than one Majorana mode at the edge (including 22111 parton order and Halperin-553 order), the situation becomes subtle. It is because the $e/4$ quasiparticles may dominate the tunneling process. In this situation, one needs to employ other types of experiments to identify different topological orders. Another tricky point for the tunneling experiment is that it may overestimate the tunneling exponent~\cite{Guang2013, Papa, Rosenow-edge, Yang-edge, Carrega-noise}. Thus the experiment provides an upper bound to the tunneling exponent. This bound may help to narrow down the set of possible candidates.

Next, the thermal Hall conductance experiment may provide a more direct probe to the topological order. If one focuses on the Pfaffian, Halperin-553, and the 22111 parton order, all of them have downstream edge modes only. Thus partial thermal equilibration should not be an issue. Lastly, all topological orders show different tunneling currents and Fano factors in the Mach-Zehnder experiment. By combining different experimental results, an unambiguous identification of the topological orders in the FQH state may be achieved.

\begin{table*}[t]
\centering
\begin{tabular}{| l | c | c | c | c | c | c | c | c | c |}
\hline
   ~~~\quad\quad\quad{Candidate} ~~~ & ~ 1C/ 2C ~ & ~~n-A?~~ 
   & ~~$g_{e/8}$~~ & ~~$g_{e/4}$~~ & ~~$\kappa_H$~~  &~even-odd effect?~ 
   & ~$(e^*/e)_{\text{max}}$~ & ~$(e^*/e)_{\text{min}}$~
    \\ \hline
    ~Pfaffian~ & ~1C~ & ~Yes~ & ~~$\bf{3/16}$~~ & ~$1/4$~ 
    & ~$3/2$~ & ~Yes~  & ~$5.63$~ & ~$0.74$~
    \\  \hline
    ~PH-Pfaffian~ & ~1C~ & ~Yes~ & ~~$\bf{3/16}$~~ & ~$1/4$~ 
    & ~$1/2$~ & ~Yes~  & ~$53.2$~ & ~$0.44$~
    \\  \hline
    ~Anti-Pfaffian/ $\bar{2}\bar{2}11111$ parton~ & ~1C~ & ~Yes~ & ~~$7/16$~~ & ~$\bf{1/4}$~ 
    & ~$1/2$~ & ~Yes~  & ~$1.93$~ & ~$0.50$~
    \\  \hline
    ~22111 parton~ & 1C & ~Yes~ & ~$7/16$~ & ~$\bf{1/4}$~ 
    & ~$5/2$~ & ~Yes~ & ~$1.08$~ & ~$0.36$~
    \\  \hline
    ~Symmetric Halperin-(5,5,3) ~ & ~2C~ & ~No~ & ~$5/16$~ & ~$\bf{1/4}$~ 
    & ~$2$~ & ~Yes~ & ~$1.93$~ & ~$0.50$~
    \\  \hline
     ~One-flavor Halperin-(5,5,3) ~ & ~2C~ & ~No~ & ~$5/16$~ & ~$\bf{1/4}$~ 
    & ~$2$~ & ~No~ & ~$2$~ & ~$43/64$~
   \\ \hline
\end{tabular}
\caption{Experimental signatures of different proposed candidates for quantum Hall state at $\nu=1/4$ in a wide quantum well. The second column classifies the topological orders into one-component (1C) or two-component (2C) states. The third column classifies the candidates into non-Abelian and Abelian orders. All listed topological orders here have a fundamental quasiparticle with charge $q=e/8$. Several other two-component candidates are discussed in Appendix~\ref{app:NA-2C} for future reference, but they are unlikely to describe the FQHE observed in the GaAs quantum well experiment. The fourth and fifth columns give the universal tunneling exponents for $e/8$ and $e/4$ quasiparticles, with the most relevant one being boldfaced. The sixth column provides the thermal Hall conductance (measured in units of $\pi^2 k_B^2 T/3h$). In the last three columns, we list the expected results from interferometry. We assume the dominant process is $e/8$ tunneling. All non-Abelian orders should demonstrate even-odd effect in a Fabry-P\'{e}rot interferometer. The Halperin-(5,5,3) order may also show the same effect, if they possess flavor symmetry. The last two columns list the maximal and minimal values of the Fano factor in a shot-noise experiment with a symmetric Mach-Zehnder interferometer  ($s=1$). }
\label{tab:summary}
\end{table*}

\section{Conclusions} \label{sec:conclusion}

To conclude our work, we have examined different experimental signatures for non-Abelian orders from composite-fermion pairing and the Abelian Halperin-(5,5,3) order for the $\nu=1/4$ FQH state. The results are summarized in Table~\ref{tab:summary}, which provides a reference and direction for future experiment to identify the underlying topological order in the system.

For the recently proposed 22111 parton order, it should show a thermal Hall conductance of 
$\kappa_H=2.5 \pi^2 k_B^2 T/3h$, satisfy the scaling laws $I\sim V^{-1/8}$ and 
$G\sim T^{-9/8}$ in tunneling experiment. In addition, we predicted that it should demonstrate a more symmetric tunneling current in the Mach-Zehnder experiment than other candidates. Furthermore, a relatively small maximal Fano factor is expected. The last two signatures provide tight constraints to test the proposal in future experiments. 

At the same time, we predicted that the two-component Halperin-(5,5,3) order should show different signatures from all the non-Abelian candidates. In particular, a measurement of $\kappa_H=2\pi^2 k_B^2 T/3h$ may be a smoking-gun signal to identify the Abelian order. Another possible way to identify the (5,5,3) order comes from Mach-Zehnder interferometry. Further support may be gained from tunneling experiment if $e/8$ quasiparticles dominate the tunneling process. More importantly, each type of experiment has its own subtleties. Therefore, the identification of the topological order in the $\nu=1/4$ FQHE requires a combination of different experimental signatures.

Lastly, some other problems on FQHE at $\nu=1/4$ are still waiting for further exploration. For example, can we have a better understanding on the phase transition in a bilayer system, in which each layer is a quarterly filled wide quantum well? Will a topological phase transition from a phase of decoupled 22111 parton orders to a high-Chern-number phase occur there? Also, what is the expected topological order in other materials with $\nu=1/4$ FQHE, such as monolayer graphene at the isospin transition point~\cite{Zibrov2018}?


\begin{acknowledgments}

The author would like to thank D. E. Feldman for his continuous guidance, insightful discussions throughout the project, and critical reading on the manuscript. Also, we would like to acknowledge M. V. Milovanovi\'{c} gratefully for valuable discussion. This work was supported by the National Science Foundation under Grant No. DMR-1607451 and the Galkin Foundation Fellowship under the Department of Physics at Brown University.
 
\end{acknowledgments}

\appendix

\section{Chern number for odd angular-momentum paired quantum Hall states}
\label{app:Chern}

In this appendix, we examine the topological nature of the non-Abelian orders originating from chiral $l$-wave pairing. Since the corresponding BCS Hamiltonian breaks both time-reversal symmetry and spin-rotational invariance, it is under the symmetry class D in the Altland-Zirnbauer classification~\cite{Altland}. The second homotopy group $\pi_2(S^2)\cong\mathbb{Z}$ suggests that the system is classified by an integer, namely the first Chern number~\cite{Read-Green}. We evaluate this quantity explicitly in the following discussion.

\begin{figure}[htb]
\begin{center}
\includegraphics[width=1.75 in]{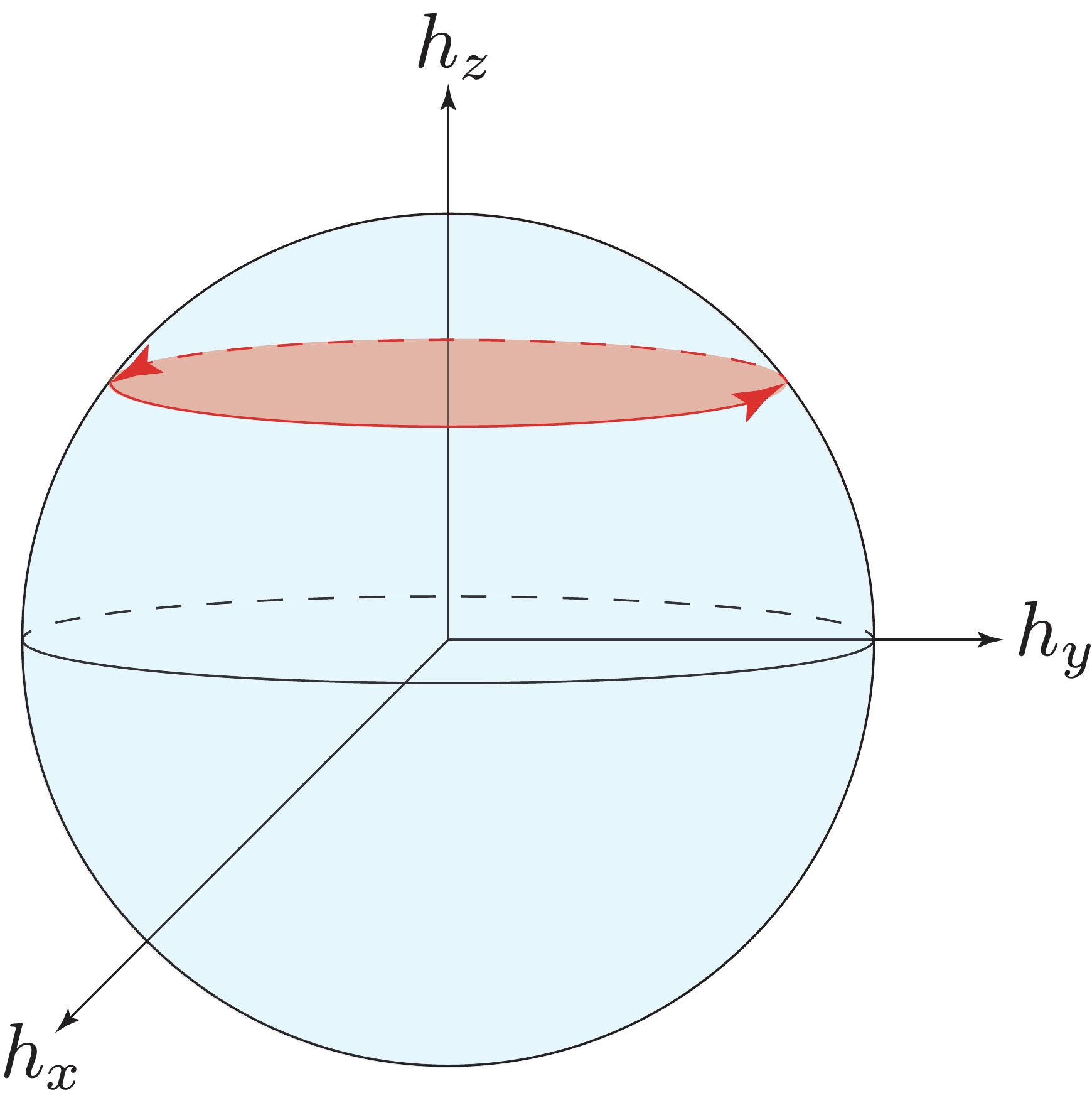}
\caption{Unit sphere $S^2$ spanned by $\hat{\bm{h}}$ when $\mu>0$. Here, we choose $\Delta_{\bm k}=\Delta_0(k_x+ik_y)^l$ as the demonstration. For a fixed value of $k$, 
$\hat{\bm{h}}$ sweeps out a circle with constant $h_z$ in the counterclockwise direction (the red circle as an example). For the $l$-wave pairing, the red circle is traversed for $l$ times. When 
$k$ increases from zero to $\infty$, $S^2$ is covered for $l$ times. Hence, the Chern number is expected to be $l$.}
\label{fig:Chern_sphere}
\end{center}
\end{figure}

In terms of field operators, the BCS Hamiltonian in Eq.~\eqref{eq:BCS_hamiltonian} in the main text can be written as
\begin{eqnarray} \label{eq:pairing_Hamiltonian}
H=\frac{1}{2}\int\frac{d^2\bm{k}}{(2\pi)^2}
\begin{bmatrix}
\Psi^{\dagger}(\bm{k}) & \Psi(-\bm{k})
\end{bmatrix}
\left(\bm{h}\cdot \bm{\sigma}\right)
\begin{bmatrix}
\Psi(\bm{k}) \\ \Psi^{\dagger}(-\bm{k})
\end{bmatrix}
\end{eqnarray}
Here, $\Psi(\bm{k})$ is the field operator which annihilates a composite fermion with momentum $\bm{k}$. The three $2\times 2$ Pauli matrices are collectively denoted as $\bm{\sigma}$. The operator $\bm{h}$ is given by
\begin{eqnarray}
\bm{h}
=\left[
\text{Re}\left(\Delta_{\bm k}\right)\quad
-\text{Im}\left(\Delta_{\bm k}\right)\quad
\frac{k^2}{2m}-\mu
\right]^T.
\end{eqnarray}
Recall that the chiral $l$-wave paired state has a gap function 
$\Delta_{\bm k}=\Delta_0(k_x\pm ik_y)^l$. In polar coordinates, one has
\begin{eqnarray} \label{eq:h-hat}
\bm{h}
=\left[\Delta_0 k^l\cos{(l\theta)}\quad
\mp \Delta_0 k^l\sin{(l\theta)}\quad
\frac{k^2}{2m}-\mu
\right]^T.
\end{eqnarray}
From this, one further defines a unit vector $\hat{\bm{h}}=\bm{h}/\left|\bm{h}\right|$. Depending on $\bm{k}$, the unit vector $\hat{\bm{h}}$ can be associated to different points on the unit sphere as shown in Fig.~\ref{fig:Chern_sphere}.

The Chern number captures the number of times that the entire unit sphere is covered when one sweeps through all possible $\bm{k}$. Explicitly, the Chern number is given by~\cite{Volovik}:
\begin{align}
\nonumber
\mathcal{C}
&=\frac{1}{4\pi}\int_{\mathbb{R}^2}
~\hat{\bm{h}}\cdot
(\partial_{k_x}\hat{\bm{h}}\times\partial_{k_y}\hat{\bm{h}})~
d^2\bm{k}
\\ \nonumber
&=\frac{1}{4\pi}\int_{\mathbb{R}^2}
~\hat{\bm{h}}\cdot
\left[
\left(\cos{\theta}\frac{\partial\hat{\bm{h}}}{\partial k}
-\frac{\sin{\theta}}{k}\frac{\partial\hat{\bm{h}}}{\partial\theta}\right)
\right.
\\
&\quad\quad\quad\quad\quad
\times
\left.
\left(\sin{\theta}\frac{\partial\hat{\bm{h}}}{\partial k}
+\frac{\cos{\theta}}{k}\frac{\partial\hat{\bm{h}}}{\partial\theta}\right)
\right]
~d^2\bm{k}
\end{align}
After a direct substitution of Eq.~\eqref{eq:h-hat}, the two-dimensional integral becomes
\begin{align}
\nonumber
\mathcal{C}
&=\mp\frac{l \Delta_0^2}{2}\int_0^\infty
\frac{k^{2l-2}\left[(l-2)\frac{k^2}{2m}-l\mu\right]}
{\left[\Delta_0 k^l)^2+(\frac{k^2}{2m}-\mu)^2\right]^{3/2}}
kdk
\\ \nonumber
&=\pm l
\left[\frac{\frac{k^2}{2m}-\mu}
{2\sqrt{\left(\Delta_0 k^l\right)^2+(\frac{k^2}{2m}-\mu)^2}}
\right]_{k=0}^{k\rightarrow +\infty}
\\ \nonumber
&=\pm\frac{l}{2}
\left[
\lim_{k\rightarrow +\infty}
\left[\left(\frac{\Delta_0 k^l}{\frac{k^2}{2m}-\mu}\right)^2+1\right]^{-1/2}
+\text{sgn}(\mu)
\right]
\\
&=\pm\frac{l}{2}
\left[1+\text{sgn}(\mu)\right].
\end{align}
In the last step, the assumption $| \Delta_0 k^l/(k^2/2m-\mu) |\ll 1$ 
for all values of $k$ has been used. Therefore, $\mathcal{C}=\pm l=\ell$ when $\mu>0$ (weak-pairing phase). On the other hand, the Chern number vanishes when $\mu<0$ (strong-pairing phase). 

\section{Wave functions for paired quantum Hall state at $\nu=1/2p$}
\label{app:wave function}

In this appendix, we introduce a class of wave functions to non-Abelian orders for the even-denominator FQH state by solving the BCS Hamiltonian in Eq.~\eqref{eq:BCS_hamiltonian}. The Hamiltonian can be diagonalized by the following Bogoliubov transformation:
\begin{eqnarray}
\begin{aligned}
b_{\bm k}
&=~u_{\bm k}c_{\bm k}-v_{\bm k}c_{-\bm k}^{\dagger},
\\
b_{\bm k}^\dagger
&=~u_{\bm k}^*c_{\bm k}^\dagger -v_{\bm k}^* c_{-\bm k},
\end{aligned}
\end{eqnarray}
where $\left|u_{\bm k}\right|^2+\left|v_{\bm k}\right|^2=1$. The diagonalized Hamiltonian takes the form
\begin{eqnarray}
H_{\rm BCS}
=\sum_{\bm k} \epsilon_{\bm k}
b_{\bm k}^\dagger b_{\bm k},
\end{eqnarray}
with the dispersion relation $\epsilon_{\bm k}=\sqrt{\xi_{\bm k}^2+\left|\Delta_{\bm k}\right|^2}$ for quasiparticle excitations. The wave function for the BCS ground state is
\begin{eqnarray}
\left|\rm BCS \right. \rangle
\sim
\prod_{\bm k}
\left|u_{\bm k}\right|^{1/2}
\exp{\left(\frac{1}{2}\sum_{\bm k}g_{\bm k}c_{\bm k}^\dagger c_{-\bm k}^\dagger\right)}
\left|\rm vac\rangle\right. .
\end{eqnarray}
The symbol $\left|\rm{vac}\rangle\right.$ denotes the vacuum state, in which no Bogoliubov quasiparticles are present. In momentum space, the correlation function $g_{\bm k}$ is given by
\begin{eqnarray} \label{eq:g_k}
g_{\bm k}
\sim \frac{1}{(k_x\pm ik_y)^l}.
\end{eqnarray}

Both the Pfaffian and PH-Pfaffian orders correspond to the paired states with $l=1$. Specifically, the former and latter have positive and negative sign in the denominator of $g_{\bm k}$, respectively. The wave function in the real space representation can be obtained from the Fourier transform of $g_{\bm k}$, i.e., $g\left(z\right)=\mathcal{F}\left[g_{\bm k}\right]$. Here, the symbol $\mathcal{F}\left[f(x)\right]$ represents the Fourier transform of the function $f(x)$. Finally, the wave function for a $\nu=1/2p$ FQH state is
\begin{eqnarray} \label{eq:wavefunction}
\Psi\left(\left\{z_i\right\}\right)
=\text{Pf}\left[g\left(z_i-z_j\right)\right]
\prod_{i<j}\left(z_i-z_j\right)^{2p}.
\end{eqnarray}
The Gaussian exponential factor has been skipped in the above equation.

To provide a demonstration of the above recipe, one can obtain the wave function for a Pfaffian state from $g\left(z\right)=\mathcal{F}[1/(k_x+ik_y)]$ as follows. To clarify notations in the following discussion, the holomorphic and antiholomorphic derivatives are defined as
\begin{eqnarray}
\frac{\partial}{\partial z}
=\frac{\partial}{\partial x}-i\frac{\partial}{\partial y}
~,~
\frac{\partial}{\partial \bar{z}}
=\frac{\partial}{\partial x}+i\frac{\partial}{\partial y}.
\end{eqnarray}
From the property of Fourier transform, one has
\begin{align}
\nonumber
&\mathcal{F}
\left[
\left(k_x+ik_y\right)\mathcal{F}^{-1}\left[g\left(z\right)\right]
\right]
\sim\delta(z)
\\ \nonumber
\Longrightarrow \quad
&\frac{\partial}{\partial \bar{z}}g\left(z\right)
\sim\delta(z)
\\ 
\Longrightarrow \quad
&g(z)\sim\frac{1}{z}=\frac{1}{x+iy}.
\end{align}
In the calculation, we only focus on the functional form in each step. Hence the symbol $\sim$ is used. All other prefactors can be absorbed in the normalization factor of the final wave function. From Eq.~\eqref{eq:wavefunction}, one obtains the wave function for the Pfaffian state ($\ell=1$ pairing):
\begin{eqnarray}
\Psi_{\rm Pf}
=
\text{Pf}\left(\frac{1}{z_i-z_j}\right)
\prod_{i<j}^N \left(z_i-z_j\right)^{2p},
\end{eqnarray}
where $N$ is the number of electrons in the system.

The above procedures can be applied to higher $l$-wave pairing, which lead to
\begin{eqnarray}
\mathcal{F}\left[\frac{1}{(k_x+ik_y)^l}\right]
\sim \frac{\bar{z}^{l-1}}{z}.
\end{eqnarray}
Thus, we obtain a possible wave function for the $\nu=1/2p$ FQHE by pairing the composite fermions in the positive $\ell$-wave channel:
\begin{eqnarray} \label{eq:l>0}
\Psi_{\ell>0}
\left(\left\{z_i\right\}\right)
=
\text{Pf}\left[\frac{\left(\bar{z}_i-\bar{z}_j\right)^{l-1}}{z_i-z_j}\right]
\prod_{i<j}^N
\left(z_i-z_j\right)^{2p}.
\end{eqnarray}

\subsection{Wave functions in the lowest Landau level}

\subsubsection{Case 1: $\ell>0$}

Clearly, the wave function in Eq.~\eqref{eq:l>0} is not confined in the lowest Landau level (LLL). To project the wave function to the LLL, one generally pulls all the antiholomorphic variables $\bar{z}_i$ to the left and replaces them by the derivatives $\bar{z}_i\rightarrow 2\frac{\partial}{\partial z_i}$~\cite{Girvin1984}. Then, the differentiation only acts on the polynomial part, but not on the exponential Gaussian factor. In our present case, this general procedure will lead to a complicated form of wave functions. Alternatively, one may obtain a possible wave function confined in the LLL by applying the procedures in Ref.~\cite{Zucker2016}:
\begin{align} \label{eq:l>0_LLL}
\nonumber
\Psi_{\ell>0}^{\rm LLL}
=
\mathcal{P}_{\rm LLL}
\left\{
\text{Pf}\left[\frac{\left(\bar{z}_i-\bar{z}_j\right)^{l-1}}{z_i-z_j}\right]
\prod_{i<j}^N
\left(z_i-z_j\right)^{2p}
\right\}.
\\
\end{align}
Here, the lowest Landau level projection operator $\mathcal{P}_{\rm LLL}$ is defined as~\cite{Zucker2016}
\begin{equation} \label{eq:P_LLL}
\mathcal{P}_{\rm LLL}
\left\{\Psi_{\ell>0}\right\}
=
\int\{d^2\xi_i\}\langle\{z_i\}|\{\xi_i\}\rangle
\Psi_{\ell>0}(\{\xi_i\}),
\end{equation}
where $\langle\{z_i\}|\{\xi_i\}\rangle=\Pi_i\exp[-(|\xi_i|^2-2\bar\xi_iz_i+|z_i|^2)/4l_B^2]$. It is plausible that $\Psi_{\ell>0}^{\rm LLL}$ is topologically equivalent to $\Psi_{\ell>0}$ in Eq.~\eqref{eq:l>0}. However, a check on whether the projected wave function truly describes a gapped phase (as required for FQH wave functions) is still lacking.

\subsubsection{Case 2: $\ell<0$}

For paired states with $\ell<0$, a possible wave function can be obtained by complex conjugating the Pfaffian factor in Eq.~\eqref{eq:l>0}. This gives
\begin{align} \label{eq:l<0}
\Psi_{\ell<0}
\left(\left\{z_i\right\}\right)
=\text{Pf}\left[
\frac{\left(z_i-z_j\right)^{| \ell |-1}}{\bar{z}_i-\bar{z}_j}
\right]
\prod_{i<j}^N \left(z_i-z_j\right)^{2p}.
\end{align}
Similar to the previous case, one may formulate a possible wave function in the LLL as
\begin{align} \label{eq:l<0_LLL}
\nonumber
\Psi_{\ell<0}^{\rm LLL}
=\mathcal{P_{\rm LLL}}
\left\{
\text{Pf}\left[
\frac{\left(z_i-z_j\right)^{| \ell |-1}}{\bar{z}_i-\bar{z}_j}
\right]
\prod_{i<j}^N \left(z_i-z_j\right)^{2p}
\right\}.
\\
\end{align}
The lowest Landau level projection is performed in the same way in Eq.~\eqref{eq:P_LLL}. Note that applying the above LLL projection to the PH-Pfaffian order leads to a gapless state, as argued in Ref.~\cite{Mishmash2018}. The same issue may happen in other negative $\ell$-wave paired states.

\subsection{How about wave functions for Abelian topological orders?}

In the above discussion, we have only formulated wave functions for non-Abelian orders by pairing the composite fermions in different odd-$l$ channels. A natural follow-up question is whether the same procedure can be applied to formulate wave functions for Abelian topological orders? For a spin-polarized or one-component system, we are not aware of how to  apply the above techniques directly on Abelian orders. Nevertheless, one can still describe the properties of the corresponding topological orders. It is believed that they are still described by the sixteenfold way~\cite{16-fold}. From this universal description, it is also possible to predict experimental signatures for them in different experiments. However, this discussion is beyond the scope of our current work and may be addressed in a separate manuscript.

At the end, we want to make a short remark which may lead to several questions for future investigation. It was shown that the wave function for spin-unpolarized Halperin-(3,3,1) order may be understood from the spin-triplet $p$-wave pairing between spin-unpolarized composite fermions~\cite{Read-Green, Ho1995}. Is it possible to generalize the idea to the $\nu=1/4$ FQHE and lead to the wave function for the Halperin-(5,5,3) order? Furthermore, the spin-polarized (3,3,1) order is related to its spin-unpolarized version by a similarity transformation~\cite{Guang2013}. Thus it leads to the following question: can wave functions for spin-polarized Abelian order can be generated indirectly from the wave functions of spin-unpolarized multicomponent Abelian orders?

\section{Other two-component candidates}
\label{app:NA-2C}

In this appendix, we provide a brief discussion of several other two-component candidates for the fractional quantum Hall state at $\nu=1/4$. For simplicity, we use the symbol $\Psi_{m,m,n}\left(\left\{z_i, w_i\right\}\right)$ to denote the wave function for the Halperin-$(m,m,n)$ order:
\begin{align}
\nonumber
&\Psi_{\left(m,m,n\right)}\left(\left\{z_i, w_i\right\}\right)
\\
=~&\prod_{i<j}\left(z_i-z_j\right)^m
\prod_{i<j}\left(w_i-w_j\right)^m
\prod_{i,j}\left(z_i-w_j\right)^n.
\end{align}
Here, $z_i$ and $w_i$ denote the coordinates for electrons in the two different layers (or pseudospins). 

\subsection{Interlayer Pfaffian order}

Similar to the $\nu=2/3$ bilayer FQH state~\cite{Peterson-2/3, Geraedts-2/3, Liu-2/3}, an interlayer Pfaffian order for the $\nu=1/4$ FQHE can be constructed. The non-Abelian order has the following wave function~\cite{wide-well, Barkeshli2010-1, Barkeshli2010-2}:
\begin{eqnarray} \label{eq:inter-Pf}
\Psi^{\text{inter}}_{(6,6,2)}
=\text{Pf}\left(\frac{1}{x_i-x_j}\right)
\Psi_{(6,6,2)}\left(\left\{z_i, w_i\right)\}\right).
\end{eqnarray}
Here, $x_i=\left\{z_i, w_i\right\}$ refers to the coordinates for all electrons in both layers. Equation \eqref{eq:inter-Pf} suggests a shift $\mathcal{S}=7$ for the topological order on a sphere. The edge consists of two Bose modes $\phi_1$, $\phi_2$ and one Majorana mode $\psi$. All modes are downstream, so the predicted thermal Hall conductance is $\kappa_H=(5/2)(\pi^2 k_B^2 T/3h)$. The wave function can be written as the following correlation function of CFT operators:
\begin{align}
\nonumber
\Psi^{\rm inter}_{(6,6,2)}
=~&\langle \mathcal{O}\left(\left\{z_i,w_i\right\}\right)\rangle
\\
=~&\Big\langle
\prod_i 
\psi(x_i) e^{6i\phi_1\left(z_i\right)}
e^{2i\phi_2\left(z_i\right)}
e^{2i\phi_1\left(w_i\right)}
e^{6i\phi_2\left(w_j\right)}
\Big\rangle.
\end{align}
We write the CFT operator for the quasiparticle as $\sigma e^{i\omega\phi_1} e^{i\eta\phi_2}$. By requiring it to have single-valued OPE with $\mathcal{O}\left(\left\{z_i, w_i\right\}\right)$, the smallest-charge quasiparticle has charge $e/8$ and is described by
\begin{eqnarray} 
\Psi_{e/8}
=\sigma e^{i\phi_1/2}e^{i\phi_2/2}.
\end{eqnarray}
The operator has scaling dimension $3/32$, so the expected tunneling exponent is $g_{e/8}=3/16$.

\subsection{Intralayer Pfaffian order}

When the interlayer correlation between electrons is stronger, one may have an intralayer Pfaffian order to describe the bilayer system~\cite{wide-well}:
\begin{align}
\nonumber
\Psi^{\text{intra}}_{(6,6,2)}
=\text{Pf}\left(\frac{1}{z_i-z_j}\right)
\text{Pf}\left(\frac{1}{w_i-w_j}\right)
\Psi_{(6,6,2)}\left(\left\{z_i, w_i\right\}\right).
\\
\end{align}
Similar to the interlayer version in Eq.~\eqref{eq:inter-Pf}, the shift for 
$\Psi_{\text{intra}}^{(6,6,2)}$ is also $\mathcal{S}=7$. However, the edge structures are different. For the intralayer version, there are two Bose modes. In addition, there are two Majorana modes, 
$\psi_1$ and $\psi_2$. Each of them is confined to a single layer. Thus, one expects to have 
$\kappa_H=3(\pi^2 k_B^2 T/3h)$. The smallest-charge quasiparticle has charge $e/16$, described by the CFT operators:
\begin{eqnarray}
\Psi_{e/16}=\sigma_1 e^{i\phi_1/2} 
\quad\text{or}\quad 
\sigma_2 e^{i\phi_2/2}
\end{eqnarray} 
Both have scaling dimension $\Delta_{e/16}=11/128$. Hence, one has $g_{e/16}=11/64$. Compared to the usual charge-$e/8$ quasiparticles in other candidates, the charge-$e/16$ quasiparticles would produce a different signature in the shot noise experiment~\cite{Heiblum-Feldman-review}.

\subsection{Singlet 22111 parton order}

The singlet 22111 parton order takes the following wave function~\cite{wide-well}:
\begin{eqnarray} \label{eq:singlet22111}
\Psi_{2_{\uparrow\downarrow}2111}
=\mathcal{P}_{\rm LLL}
\left[\chi_1(\left\{z_i\right\}) \chi_1(\left\{w_i\right\}) \chi_2 \chi_1^3\right].
\end{eqnarray}
Here, $\chi_n$ denotes the wave function for the integer quantum Hall state with $n$ completely filled Landau levels. The parton order has a shift $\mathcal{S}=6$, which is different from the fully spin-polarized version in the main text. From a similar discussion on 
$2_{\uparrow\downarrow}21$ parton order at $\nu=1/2$~\cite{Read-Green, Milovanovic-d-wave}, it is believed that the $2_{\uparrow\downarrow}2111$ parton order is Abelian and can be understood as the result of a $d$-wave pairing between composite fermions. The edge structure of the parton order consists of two Bose modes, so the predicted thermal Hall conductance is $\kappa_H=2(\pi^2 k_B^2 T/3h)$. The fundamental quasiparticle has charge $e/8$ with a scaling dimension $\Delta_{e/8}=5/32$. This leads to a predicted tunneling exponent $g_{e/8}=5/16$.

\subsection{Shift and predicted experimental signatures}

In Table~\ref{tab:summary-2C}, some predicted experimental signatures and shifts for the two-component topological orders are summarized. A combination of a tunneling experiment and a thermal conductance experiment can distinguish the two-component orders here and the topological orders in the main text.

\begin{table} [htb]
\begin{tabular}  {| l | c | c |c | c | c | }
\hline
   \quad{Candidate} ~~ & ~~n-A?~~  & ~~$Q_{\rm qp }$~~ & ~~$g_{qp}$~~ 
   & ~~$\kappa_H$~~ & ~~$\mathcal{S}$~~
    \\ \hline
    ~$\Psi_{(5,5,3)}$ ~ & ~No~ & ~$e/8$~ & ~$5/16$~ & ~$2$~ & ~$5$~
    \\ \hline
     ~$\Psi^{\rm inter}_{(6,6,2)}$ ~ & ~Yes ~ & ~$\bm{e/8}$~ & ~$\bm{3/16}$~  & ~$5/2$~ & ~$7$~
   \\ \hline
    ~$\Psi^{\rm intra}_{(6,6,2)}$ ~ & ~Yes ~ & ~$\bm{e/16}$~ & ~$\bm{11/64}$~  & ~$3$~ & ~$7$~
     \\ \hline
    ~$\Psi_{2_{\uparrow\downarrow}2111}$~ & ~No~ & ~$e/8$~ & ~$5/16$~ & ~$2$~  & ~$6$~ 
   \\ \hline
\end{tabular}
\caption{Predicted experimental signatures and shifts for different two-component candidates for FQHE at $\nu=1/4$. The second column classifies the candidates into non-Abelian and Abelian orders. The third and fourth columns list the charges of fundamental quasiparticles and their corresponding tunneling exponents. The values are boldfaced if the quasiparticles are the  most relevant in the topological order. The fifth column gives the predicted thermal Hall conductances, in units of $\pi^2 k_B^2 T/3h$. Note that all candidates here have downstream edge modes only. The last column shows the shift of the topological order. }
\label{tab:summary-2C}
\end{table}


\begin{thebibliography}{99}
\bibitem{foot-335}
The direct analogy of Halperin-113 order for $\nu=1/4$ FQH state should be Halperin-335 order. Nevertheless, we do not discuss it in the current manuscript.
\end{thebibliography}
\end{document}